\documentclass[reprint,nofootinbib,aps]{revtex4-1}
\usepackage{amsmath,amssymb,amsbsy,amstext,amsthm,simplewick,amsfonts} 
\usepackage{physics}
\usepackage{graphicx} 
\usepackage{dcolumn} 
\usepackage{bm} 
\usepackage{comment}
\usepackage{enumitem}
\usepackage{hyperref}
\hypersetup{
	colorlinks = true,
	linkcolor = blue,
	citecolor = blue,
	urlcolor = blue
}

\begin{document}

\title{Axion dark matter from inflation-driven quantum phase transition}
\author{Ameen Ismail$^{1}$}
\email{ai279@cornell.edu}

\author{Seung J. Lee$^{2}$}
\email{sjjlee@korea.ac.kr}

\author{Bingrong Yu$^{1,2}$}
\email{bingrong.yu@cornell.edu}

\affiliation{
	$^1$ Department of Physics, LEPP, Cornell University, Ithaca, NY 14853, USA\\
	$^2$ Department of Physics, Korea University, Seoul 136-713, Korea
}

\date{\today}

\begin{abstract}
We propose a new mechanism to produce axion dark matter from inflationary fluctuations. 
Quantum fluctuations during inflation are strengthened by a coupling of the axion kinetic term to the inflaton, which we parametrize as an effective curvature $\kappa$ in the axion equation of motion. A nonvanishing curvature breaks the scale invariance of the axion power spectrum, driving a quantum phase transition with $\kappa$ as the order parameter. The axion power spectrum is proportional to the inverse comoving horizon to the power of $\kappa$. For positive $\kappa$ the spectrum gets a red tilt, leading to an exponential enhancement of the axion abundance as the comoving horizon shrinks during inflation. This enhancement allows sufficient axion production to comprise the entire dark matter relic abundance despite the ultralight mass. 
Our mechanism predicts a significantly different parameter space from the usual misalignment mechanism. It allows for axion-like particle dark matter with a much lower decay constant and thus a larger coupling to Standard Model particles. Much of the parameter space can be probed by future experiments including haloscopes, nuclear clocks, CASPEr, and CMB-S4. We can also generate heavier QCD axion dark matter than the misalignment mechanism.
\end{abstract}

\maketitle

\section{Introduction}

Dark matter (DM) appears to be an indispensable component of modern cosmology. In recent years, there has been a growing interest in ultralight DM, i.e., DM particles with sub-eV masses~\cite{Ferreira:2020fam,Hui:2021tkt,Antypas:2022asj}. At such a low mass, DM exhibits wave-like properties because its de Broglie wavelength exceeds the interparticle separation, given the DM density in the solar neighborhood $\rho_{\rm DM}\approx 0.4~{\rm GeV}/{\rm cm}^3$. Therefore, ultralight DM can be probed by astrophysics and atomic physics experiments that exploit its oscillatory and interfering behaviors. Typical examples include pulsar timing arrays~\cite{Kim:2023kyy} and quantum sensors~\cite{Kim:2022ype,Flambaum:2023bnw,Flambaum:2023drb,Zhao:2024tvg}. In particular, future atomic/nuclear clocks will be able to probe ultralight DM masses down to $10^{-22}~{\rm eV}$~\cite{Kim:2022ype,Flambaum:2023bnw}. This already enters the range of fuzzy DM~\cite{Hu:2000ke}, which can help resolve the small-scale crisis of cold DM~\cite{Hui:2016ltb}, and is also preferred by a recent analysis of gravitational lens data~\cite{Amruth:2023xqj} .

The quantum chromodynamics (QCD) axion, as well as axion-like particles (ALPs), are well-motivated candidates for ultralight DM. Axions are pseudo-Nambu--Goldstone bosons (pNGBs) of a spontaneously broken  $U(1)$ Peccei--Quinn (PQ) symmetry~\cite{Weinberg:1977ma,Wilczek:1977pj}. The QCD axion provides a solution to the strong CP problem through its coupling to the gluon~\cite{Peccei:1977hh,Peccei:1977ur}. The axion mass is protected from large radiative corrections due to the shift symmetry and its interactions are suppressed by the PQ breaking scale. This renders the axion a light and long-lived particle, and thus suitable to serve as ultralight DM (for recent reviews see Refs.~\cite{Marsh:2015xka,DiLuzio:2020wdo}).

The usual misalignment mechanism for producing the axion relic abundance predicts a relationship between the axion mass and the PQ breaking scale~\cite{Dine:1982ah,Preskill:1982cy,Arias:2012az}. Assuming that the axion comprises all of the DM, this restricts the viable parameter space. 
For instance, for QCD axion DM, the mass should be of order $10^{-6}~{\rm eV}$ (assuming the initial misalignment angle is not fine-tuned)~\cite{GrillidiCortona:2015jxo}. On the other hand, if we consider ALP DM lighter than $10^{-12}~{\rm eV}$, which is the regime where future quantum sensor probes are most sensitive~\cite{Kim:2022ype,Flambaum:2023bnw,Flambaum:2023drb,Zhao:2024tvg}, then the PQ breaking scale has to be higher than $10^{14}~{\rm GeV}$. Consequently, the couplings to Standard Model (SM) particles are highly suppressed and difficult to probe experimentally. This has motivated the development of new mechanisms for producing axion DM that allow for a different relationship between mass and coupling, such as parametric resonance~\cite{Co:2017mop}, nonstandard cosmologies~\cite{Blinov:2019rhb}, kinetic misalignment~\cite{Co:2019jts}, ALP cogenesis~\cite{Co:2020xlh}, and trapped misalignment~\cite{DiLuzio:2021gos} (see~\cite{Takahashi:2018tdu,Co:2018mho,Ho:2019ayl,Harigaya:2019qnl,Asadi:2022njl,Kitano:2023mra} for more possibilities).
 
In this work, we propose a new mechanism to produce axion DM from inflationary fluctuations, which allows for a much lower breaking scale than misalignment. The basic idea is to make use of quantum fluctuations during the cosmic inflationary epoch to efficiently produce the axion abundance. Typically, inflationary fluctuations do not lead to a sufficient DM production, due to the smallness of the axion mass~\cite{Graham:2015rva,Redi:2022llj}. However, if the axion kinetic term is coupled to the inflaton, the relic abundance can be exponentially enhanced\footnote{A similar idea was first raised in Ref.~\cite{Nakai:2020cfw} to produce dark photon DM through exponential enhancement, where a specific form of kinetic coupling was used.}. The kinetic coupling breaks the scale-invariant power spectrum of a massless axion and, together with the quantum fluctuations during inflation, drives a ``quantum phase transition'' from the conformal symmetric phase to the broken phase. We will show how one can calculate the tilt in the power spectrum from the form of the kinetic coupling. We find that if the power spectrum is broken to a red tilt, the axion abundance gets exponentially enhanced by the number of inflationary e-folds, compensating the suppression of the DM density by the small axion mass.

Our mechanism is applicable to both the QCD axion and ALPs. The relic density does not depend on the PQ breaking scale, which opens up a large window in the axion DM parameter space across a wide range of masses down to $\sim 10^{-24}~{\rm eV}$. The axion couplings can be much larger than one obtains in the misalignment mechanism. Much of the parameter space can be probed by future axion DM experiments such as haloscopes, nuclear clocks, and CASPEr (see Fig.~\ref{fig:photongluon coupling}). Our mechanism can also potentially be probed by next-generation cosmic microwave background (CMB) experiments and $21$~cm telescopes.

\section{Quantum phase transition driven by inflation}
\subsection{Framework}
We start from the following action:
\begin{align}
	\label{eq:action}
	{\cal S} = \int {\rm d}^4 x \sqrt{-g}&\left[\frac{M_{\rm Pl}^2}{2}R-\frac{1}{2}g^{\mu\nu}\partial_\mu \phi \partial_\nu \phi-V(\phi)\right.\nonumber\\
	&\left.-\frac{1}{2}K^2(\phi)g^{\mu\nu}\partial_\mu \eta \partial_\nu \eta-\frac{1}{2} m_\eta^2 \eta^2\right]\;,
\end{align} 
where $M_{\rm Pl}$ is the reduced Planck mass, $R$ is the Ricci scalar, and $V(\phi)$ is the inflaton potential. The axion 
$\eta$ is coupled to the inflaton $\phi$ through its kinetic term during inflation, with $K(\phi)$ an arbitrary function of the inflaton. Such a noncanonical scalar kinetic term can be naturally realized, e.g., in the framework of supergravity~\cite{Ellis:1983sf,Ellis:2013xoa} with the scalar $\eta$ identified as the modulus field coming from orbifold compactifications~\cite{Kobayashi:2016mzg,Abe:2023ylh}. An example of PQ unification is given in Appendix~\ref{app:examples}, where a nonminimal kinetic coupling naturally arises.
The kinetic function $K(\phi)$ should smoothly reduce to unity at the end of inflation, such that the axion is decoupled from the inflaton thereafter and we are left with a canonical kinetic term. We take the usual flat FLRW metric $g_{\rm \mu\nu}=a^2(\tau){\rm diag}\left(-1,1,1,1\right)$, where $a$ is the scale factor and $\tau$ is the conformal time. 

The axion field $\eta$ is the angular mode of a complex scalar $\chi=\rho\,e^{{\rm i}\eta/f_\eta}/\sqrt{2}$, and appears as a Nambu--Goldstone boson after the spontaneous breaking of PQ symmetry. A nonzero axion mass $m_\eta$ can come from nonperturbative effects which explicitly break the shift symmetry. In this work, we do not specify the origin of the axion mass and treat it as a free parameter. In Eq.~(\ref{eq:action}), we have assumed $f_\eta > H_{\rm inf}/2\pi$, where $H_{\rm inf}$ is the Hubble parameter during inflation. Thus the PQ symmetry is broken and the axion is a physical degree of freedom during inflation. The radial mode $\rho$ is not included in Eq.~(\ref{eq:action}) since its mass scale $m_\rho= {\cal O}(f_\eta)$ is higher than the inflation scale and is assumed to be integrated out.  

From the action in Eq.~(\ref{eq:action}), one obtains the equation of motion (EOM) for the axion: 
\begin{align}
	\label{eq:EOM}
	f'' - \nabla^2 f - \left(\frac{a''}{a}+\frac{K''}{K}+2\frac{a'}{a}\frac{K'}{K}+\frac{a^2 m_\eta^2}{K^2}\right)f=0\;,
\end{align}
where $f\equiv aK\eta$ and the primes denote derivatives with respect to the conformal time $\tau$. To simplify the EOM, we define the following dimensionless quantities:
\begin{align}
	\label{eq:curvaturedef}
	\kappa_1 \equiv \tau^2 \frac{K''}{K^2}\;,\quad
	\kappa_2 \equiv -\tau\frac{K'}{K}\;, \quad
	\kappa\equiv \kappa_1 + 2\kappa_2\;.
\end{align}
Under the slow-roll approximation, $\kappa_1$ and $\kappa_2$ can be directly computed from the kinetic function $K(\phi)$ and inflaton potential $V(\phi)$:
\begin{align}
	\label{eq:curvature}
	\kappa_1 &\approx M_{\rm Pl}^2\left(2\epsilon\frac{K_{\phi\phi}}{K}-\frac{K_\phi}{K}\frac{V_\phi}{V}\right),\;
	\kappa_2 \approx -M_{\rm Pl}^2 \frac{K_\phi}{K}\frac{V_\phi}{V},
\end{align}
where $\epsilon \equiv M_{\rm Pl}^2\left(V_\phi/V\right)^2/2 \ll 1$ is the first slow-roll parameter and the subscript $\phi$ denotes the derivative with respect to $\phi$.  So up to ${\cal O}(\epsilon)$ corrections, we have $\kappa \approx 3\kappa_1$. Since $\kappa >0 $ corresponds to the geometric picture where the curve $K(\tau)$ has a positive curvature during inflation, we call $\kappa$ \emph{effective curvature} from now on. As we will see later, a positive $\kappa$ can lead to an exponential enhancement of the axion abundance\footnote{To make our discussion more general, we do not specify the explicit form of kinetic coupling. Some concrete examples are given in Appendix~\ref{app:examples}.}.

During inflation, the spacetime is nearly de Sitter, which means the Hubble parameter is approximately constant and we have $a\approx -1/(\tau H_{\rm inf})$. Then the EOM is reduced to
\begin{align}
	\label{eq:EOM-k}
	f_k '' + \left(k^2 - \frac{2+\kappa}{\tau^2}\right)f_k=0\;,
\end{align}
where $f_k$ is the Fourier mode of $f$ with comoving momentum $k$. In Eq.~(\ref{eq:EOM-k}), we have neglected the axion mass term, which is reasonable if $\left|m_\eta/(K H_{\rm inf})\right| \ll 1$ is satisfied. In general, the effective curvature $\kappa$ depends on $\phi$, but as long as $\kappa$ varies slowly during inflation compared with the axion field $f_k$, we can treat it as a constant.

Taking the Bunch--Davies initial condition, the solution of Eq.~(\ref{eq:EOM-k}) is given by 
\begin{align}
 	f_k (\tau) = \frac{\sqrt{\pi}}{2}\sqrt{-\tau}H^{(1)}_{\nu}\left(-k\tau\right)\;,\quad \nu\equiv \sqrt{9/4+\kappa}\;,
\end{align}
where $H_\nu^{(1)}$ is the Hankel function of the first kind. 

\subsection{Breaking conformal symmetry}
The effective curvature $\kappa$ plays a crucial role in producing the axion during inflation. To see this more clearly, we compute the axion power spectrum $P_k$, which is proportional to the two-point correlation function $\left\langle f_k^{} f_k^* \right\rangle$ generated by quantum fluctuations. For superhorizon modes (i.e. $x\equiv -k\tau < 1$), we find:
\begin{align}
	\label{eq:powerspectrum}
	P_k  \sim \left(\frac{H_{\rm inf}}{2\pi}\right)^2 \left(-k\tau\right)^{3-2\nu}\sim \left(\frac{H_{\rm inf}}{2\pi}\right)^2 \left(\frac{1}{x}\right)^{2\kappa/3}\;.
\end{align} 
Here $H_{\rm inf}/2\pi$ is the Gibbons--Hawking temperature~\cite{Gibbons:1977mu} that characterizes the typical magnitude of quantum fluctuations during inflation. In the last step of Eq.~(\ref{eq:powerspectrum}), we have assumed $\left|\kappa\right|<1$ due to the backreaction constraint (see below).

At the critical point $\kappa=0$, the power spectrum is scale-invariant, reflecting the conformal symmetry of a free massless scalar. However, a nonzero $\kappa$ breaks the scale invariance of the spectrum. This may remind one of the anomalous dimension (given by $\gamma\equiv 2\kappa/3$) that breaks the classical scaling behavior of the correlation function. On the other hand, $\kappa$ also plays the role of an \emph{order parameter} that drives a phase transition from the conformal symmetric phase to the broken phase. Since the axion has no physical vacuum expectation value during inflation due to the lack of an axion potential to break shift symmetry\footnote{The axion mass does break shift symmetry, but the mass effect is negligible during inflation for an ultralight axion.}, quantum fluctuations are the only source to produce the axion abundance. This is in contrast to a classical phase transition driven by thermal fluctuations. Moreover, the order parameter $\kappa$ comes from kinetic coupling, which may have an origin of quantum corrections from couplings to higher-order operators (see Appendix~\ref{app:examples}). The above features make it a \emph{quantum phase transition}.

\begin{figure}[t!]
	\centering
	\includegraphics[width=0.98\linewidth]{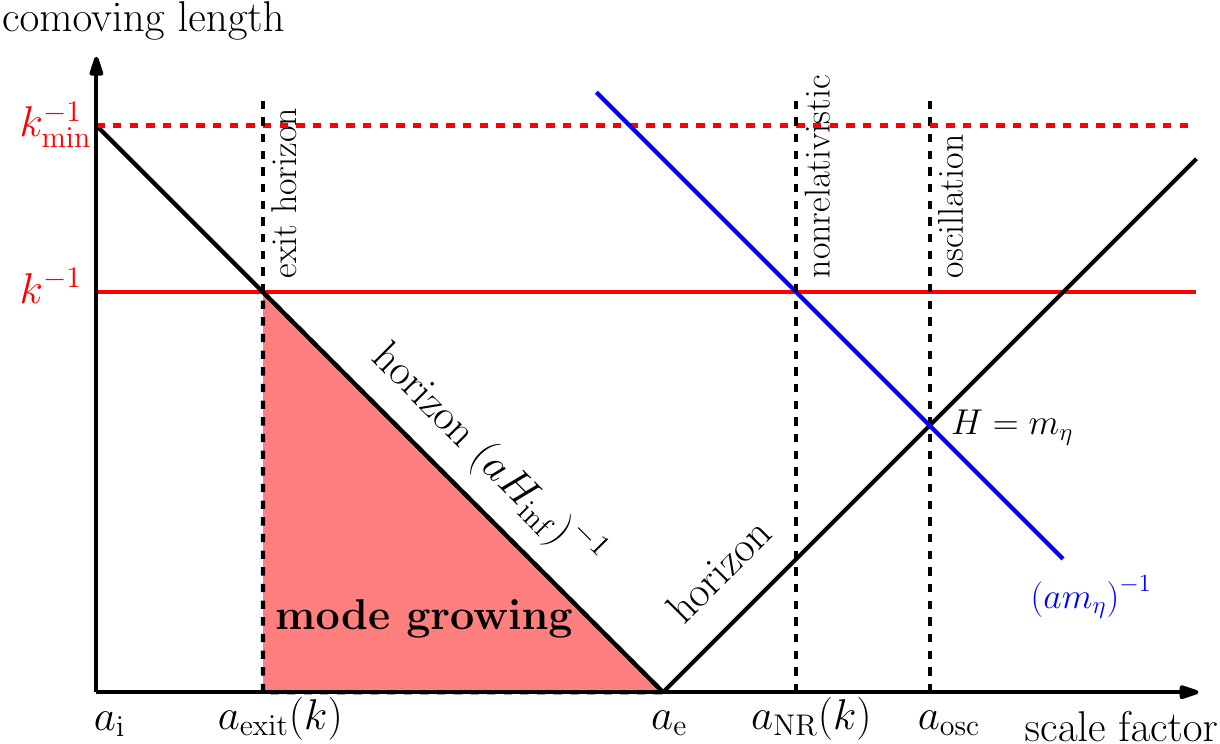}
	\caption{\label{fig:horizon} A sketch of our mechanism. The comoving horizon shrinks during inflation and increases later. For each mode $k$, with a positive effective curvature, it starts to grow after exiting the horizon (red shaded region). The minimum mode $k_{\rm min}\propto a_{\rm i}H_{\rm inf}$ (red dashed line) is set by the initial horizon, and receives largest enhancement during inflation. The comoving Compton wavelength $1/(a m_\eta)$ is shown with blue line for a fixed axion mass $m_\eta$. The mode $k$ becomes nonrelativistic after $a_{\rm NR}(k)=k/m_\eta$ and begins to oscillate when $a>a_{\rm osc}$ with $a_{\rm osc}$ determined by $H(a_{\rm osc})=m_\eta$. Unlike misalignment, our mechanism allows axion to become nonrelativistic well before coherent oscillation, i.e., $a_{\rm NR}\ll a_{\rm osc}$.}  
\end{figure}

In particular, a positive $\kappa$ leads to a red tilt, with the axion power spectrum dominated by superhorizon modes. Note that in Eq.~(\ref{eq:powerspectrum}), we have $1/x=a H_{\rm inf}/k$, which means the power spectrum obeys a power law proportional to the inverse comoving horizon $a H_{\rm inf}$ to the power of $\kappa$.
Since the comoving horizon is exponentially shrinking during inflation, the power spectrum will get exponentially enhanced if $\kappa>0$. For each mode $k$, it starts to grow only after exiting the horizon, i.e., $\left(a H_{\rm inf}\right)^{-1}<k^{-1}$. This corresponds to the shaded region in Fig.~\ref{fig:horizon}. Modes with smaller $k$ get more enhancement during inflation (larger shaded area).
On the other hand, as $\kappa$ goes from zero to negative, the power spectrum gets a blue tilt, scaling as $(k/(aH_{\rm inf}))^{-2\kappa/3}$, and then all superhorizon modes are suppressed by the end of inflation.

We can verify this by calculating the axion energy density from the energy-momentum tensor. 
For superhorizon modes, one can use the small-value expansion of Hankel functions and keep only the leading term 
\begin{align}
	\label{eq:energy}
	\langle \rho_\eta (\tau) \rangle \approx &\frac{H_{\rm inf}^4}{16\pi^3}\,2^{2\nu} \left(\frac{\kappa}{3}+\frac{3}{2}-\nu\right)^2\Gamma^2(\nu)\nonumber\\&\times  \int_{-k_{\rm min}\tau}^{{\cal O}(1)} \frac{{\rm d}x}{x} \left(\frac{1}{x}\right)^{2\nu-3}\;,
\end{align}
where the angle bracket denotes the average over different modes and the upper limit of the integral is some ${\cal O}(1)$ number. As we can see from Eq.~(\ref{eq:energy}), for a positive effective curvature ($\nu>3/2$), the spectrum is indeed red. In this case, the upper limit of the integral in (\ref{eq:energy}) does not affect the result; the energy density is dominated by the minimum mode $k_{\rm min}$ (see the red dashed line in Fig.~\ref{fig:horizon}):
\begin{align}
	k_{\rm min} = \sqrt{\kappa}a_{\rm i}H_{\rm inf} = -\sqrt{\kappa}/\tau_{\rm i}\;,
\end{align}
where $a_{\rm i}$ and $\tau_{\rm i}$ are the scale factor and conformal time at the beginning of inflation, respectively. This mode exits the horizon at $\tau_{\rm i}$ and receives the largest exponential enhancement during inflation. Modes with comoving momentum smaller than $k_{\rm min}$ do not enter the horizon at any time and do not contribute to physical observables. 
Working out the integral, we obtain the energy density at the end of inflation:
\begin{align}
	\label{eq:enerngy at ae}
	\langle \rho_\eta (\tau_{\rm e}) \rangle \approx \frac{H_{\rm inf}^4}{16\pi^3}\,{\cal F}(\kappa)\,e^{N\left(2\nu-3\right)}\;,
\end{align}
where
\begin{align}
{\cal F}(\kappa)\equiv\frac{2^{2\nu}\left(\kappa/3+3/2-\nu\right)^2 \Gamma^2(\nu)}{2\nu - 3}\left(\frac{1}{\kappa}\right)^{\nu-3/2}\;.
\end{align}
Here we have used the relation $\tau_{\rm i}/\tau_{\rm e}=e^N$, with $N$ the number of inflationary e-folds. Now it is clear that the exponential enhancement of the energy density comes from the large hierarchy between the comoving horizon at the beginning and the end of inflation. The magnitude of the enhancement is modulated by the effective curvature $\kappa$ which breaks the conformal invariance of a massless axion. In particular, for $0<\kappa\ll 1$, we have
\begin{align}
	\langle \rho_\eta (\tau_{\rm e}) \rangle \approx \frac{H_{\rm inf}^4}{3888\pi^2}\kappa^3\, e^{2\kappa N/3}\;.
\end{align}
Note that as $\kappa \to 0$, the scale-invariant spectrum is restored, and the contributions from modes other than $k_{\rm min}$ to the energy density are not negligible.

\subsection{Backreaction constraint}
In principle, the kinetic coupling between the axion and the inflaton would also affect the dynamics of the inflaton. However, we want the backreaction of the axion on the inflaton to be small enough that the formalism of single-field inflation still holds. 

From Eq.~(\ref{eq:action}), the EOM of the inflaton is given by
\begin{align}
	\ddot{\phi} + 3H_{\rm inf} \dot{\phi} + V_\phi + K K_\phi g^{\mu\nu}\partial_\mu\eta \partial_\nu\eta=0\;,
\end{align}
where the dot denotes the derivative with respect to the physical time $t$. In order to not affect the inflaton dynamics, we require
\begin{align}
	\label{eq:backreaction1}
	\left|K K_\phi \langle g^{\mu\nu}\partial_\mu\eta \partial_\nu\eta\rangle \right| \ll \left|3H_{\rm inf} \dot{\phi} \right|\;.
\end{align}
In addition, the axion energy density produced during inflation should be negligible compared to the  energy density of the inflaton, so 
\begin{align}
	\label{eq:backreaction2}
	\langle \rho_\eta \rangle \ll 3 M_{\rm Pl}^2 H_{\rm inf} ^2\;.
\end{align}
Combining Eqs.~(\ref{eq:backreaction1}) and (\ref{eq:backreaction2}), we obtain the following backreaction constraint on the effective curvature $\kappa$:
\begin{align}
	\label{eq:backreaction}
\kappa\,{\cal F}(\kappa)\,e^{N\left(2\nu-3\right)} \ll 18\pi/A_{\rm s}\;,
\end{align}
where $A_{\rm s}\equiv H_{\rm inf}^2/\left(8\pi^2\epsilon M_{\rm Pl}^2\right) = 2.2 \times 10^{-9}$ is the scalar amplitude measured at the pivot scale $k_*\equiv 0.05~{\rm Mpc}^{-1}$~\cite{Planck:2018vyg}. Since the current Planck measurement on the curvature perturbation is at the percent level~\cite{Planck:2018vyg}, we conservatively require that the backreaction effect of the axion on the inflaton is less than one percent.
This puts a strict upper bound on the magnitude of scale invariance breaking. For example, from Eq.~(\ref{eq:backreaction}), for $N=50$, $N=60$ and $N=70$, we obtain $\kappa<0.79$, $\kappa<0.67$ and $\kappa<0.58$, respectively. 


\section{Relic abundance}
The axion produced during inflation has a spectrum peaked at the superhorizon mode, which leaves a homogeneous background from the view of CMB modes~\cite{Nakai:2020cfw,Bartolo:2012sd}. In addition, as we will show below, the axion can become nonrelativistic well before structure formation, and therefore serves as a good candidate for cold DM.

The physical momentum of the axion at the end of inflation is peaked at
\begin{align}
	p_{\rm e} = k_{\rm min}/a_{\rm e}=\sqrt{\kappa} e^{-N} H_{\rm inf}\;.
\end{align}
The following cosmological evolution of the axion abundance depends on the relative size of $p_{\rm e}$ and $m_\eta$. We will first discuss the case of ALPs, then the QCD axion.

\begin{figure}[t!]
	\centering
	\includegraphics[width=0.98\linewidth]{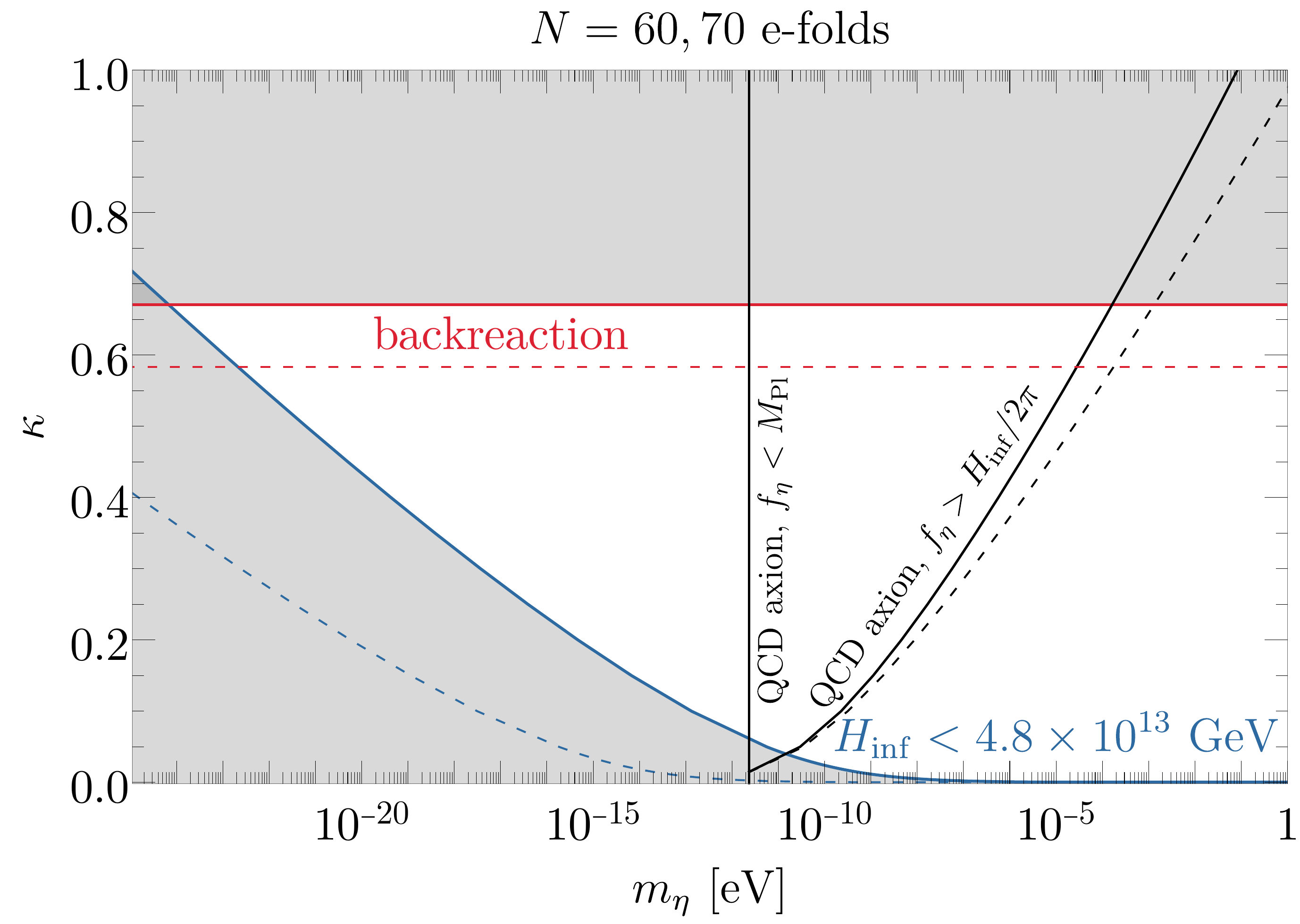}
	\caption{\label{fig:parameter}Parameter space of the axion mass $m_\eta$ and effective curvature of the inflaton-axion coupling $\kappa$ that predict the correct DM abundance in our mechanism, with a fixed number of e-folds $N=60$ (solid lines) or $N=70$ (dashed lines). The red lines denote the backreaction constraint in Eq.~\eqref{eq:backreaction}. The blue lines correspond to the upper bound on the inflationary Hubble scale ($H_{\rm inf} < 4.8 \times 10^{13}$~GeV), which is constrained by the tensor-to-scalar ratio. For the QCD axion, the parameter space is further restricted between the two black lines from the requirements that the decay constant be sub-Planckian ($f_\eta < M_{\rm Pl}$) and that PQ symmetry is broken during inflation ($f_\eta > H_{\rm inf}/2\pi$).}  
\end{figure}

\subsection{Axion-like particles}
For ALPs, we do not specify the origin of $m_\eta$ and only treat it as a free parameter during inflation. If $m_\eta<p_{\rm e}$, the axion is relativistic at the end of inflation. Later, it becomes nonrelativistic due to redshift at temperature $T_{\rm NR}$, which is determined by
\begin{align}
	\label{eq:TNR}
	T_{\rm NR}=\frac{m_\eta} {p_{\rm e}} T_{\rm reh}=\frac{1}{\sqrt{\kappa}}\frac{m_\eta}{H_{\rm inf}}T_{\rm reh}e^N\;,
\end{align}
where $T_{\rm reh}$ is the reheating temperature, and for simplicity we assume an instantaneous reheating after inflation. Due to the $e^N$ enhancement on the right-hand side of Eq.~(\ref{eq:TNR}), it is easy for the axion to become nonrelativistic before structure formation, $T\sim {\rm keV}$. Therefore, unlike
misalignment, our mechanism allows axion to become nonrelativistic much earlier than it begins to oscillate, as can be seen clearly from Fig.~\ref{fig:horizon}.

The current axion energy density is connected to that at the end of inflation by $\langle \rho_\eta(\tau_0) \rangle= \langle \rho_\eta(\tau_{\rm e}) \rangle\left(a_{\rm e}/a_{\rm NR}\right)^4\left(a_{\rm NR}/{a_{0}}\right)^3$, where $a_{\rm NR}$ is the scale factor at $T_{\rm NR}$. Assuming a standard cosmology and using Eq.~(\ref{eq:enerngy at ae}), one obtains the axion relic abundance $\Omega_\eta  \equiv \langle \rho_\eta(\tau_0) \rangle/\rho_{\rm c}$ at the present day:
\begin{eqnarray}
	\label{eq:abundance-relativistic}
	\Omega_\eta
	=\frac{g_{*0}g_{*\rm reh}^{-1/4}}{48\pi^3}\left(\frac{\pi^2}{90}\right)^{3/4}\frac{m_\eta T_0^3  H_{\rm inf}^{3/2}}{M_{\rm Pl}^{7/2}H_0^2}\frac{{\cal F}(\kappa)}{\sqrt{\kappa}}e^{N\left(2\nu-2\right)},
\end{eqnarray}
where $\rho_{\rm c}=3H_0^2 M_{\rm Pl}^2$ is the critical density with $H_0$ the current Hubble parameter, and $T_0=2.725~{\rm K}$ is the current cosmic temperature. In addition, $g_{*\rm reh}=106.75$ and $g_{*\rm 0}=2$ are the relativistic degrees of freedom at the reheating epoch and present day. We emphasize that the small axion mass in Eq.~(\ref{eq:abundance-relativistic}) is compensated by the exponential enhancement driven by the effective curvature, such that $\Omega_\eta$ can in principle comprise all of the DM.

On the other hand, a heavier axion satisfying  $m_\eta > p_{\rm e}$ is already nonrelativistic at the end of inflation. So the present-day energy density is given by $\langle \rho_\eta(\tau_0) \rangle= \langle \rho_\eta(\tau_{\rm e}) \rangle\left(a_{\rm e}/a_{0}\right)^3$ and we obtain
\begin{align}
	\label{eq:abundance-NR}
	\Omega_\eta=\frac{g_{*0}g_{*\rm reh}^{-1/4}}{48\pi^3}\left(\frac{\pi^2}{90}\right)^{3/4}\frac{T_0^3 H_{\rm inf}^{5/2}}{M_{\rm Pl}^{7/2}H_0^2}{\cal F}(\kappa)\,e^{N\left(2\nu-3\right)}\;.
\end{align}
Note that in this case, the relic abundance does not depend on the axion mass, and it corresponds to a scenario with a much lower inflationary Hubble scale $H_{\rm inf}$.

The viable parameter space for our axion DM is shown in Fig.~\ref{fig:parameter}, where we present the backreaction constraint Eq.~\eqref{eq:backreaction} and the upper bound on the inflationary Hubble scale $H_{\rm inf} < 4.8 \times 10^{13}$~GeV. The upper bound on $H_{\rm inf}$ comes from the constraint on the tensor-to-scalar ratio $r_{\rm T}<0.036$~\cite{BICEP:2021xfz}, using the relation $H_{\rm inf}/2\pi = M_{\rm Pl}\sqrt{A_{\rm s}r_{\rm T}/8}$. For $N=60$ e-folds of inflation, we can achieve an axion as light as $10^{-24}~{\rm eV}$. A larger number of e-folds leads to a greater exponential enhancement of the axion abundance (though the backreaction constraint on $\kappa$ is stricter), and thus allows a lighter axion. For instance, for $N=70$, the minimum axion mass could reach even $10^{-29}~{\rm eV}$.
There are also phenomenological bounds on the minimum DM mass, of course, but we will discuss these in the next section.

\begin{figure*}
	\includegraphics[width=0.48\textwidth]{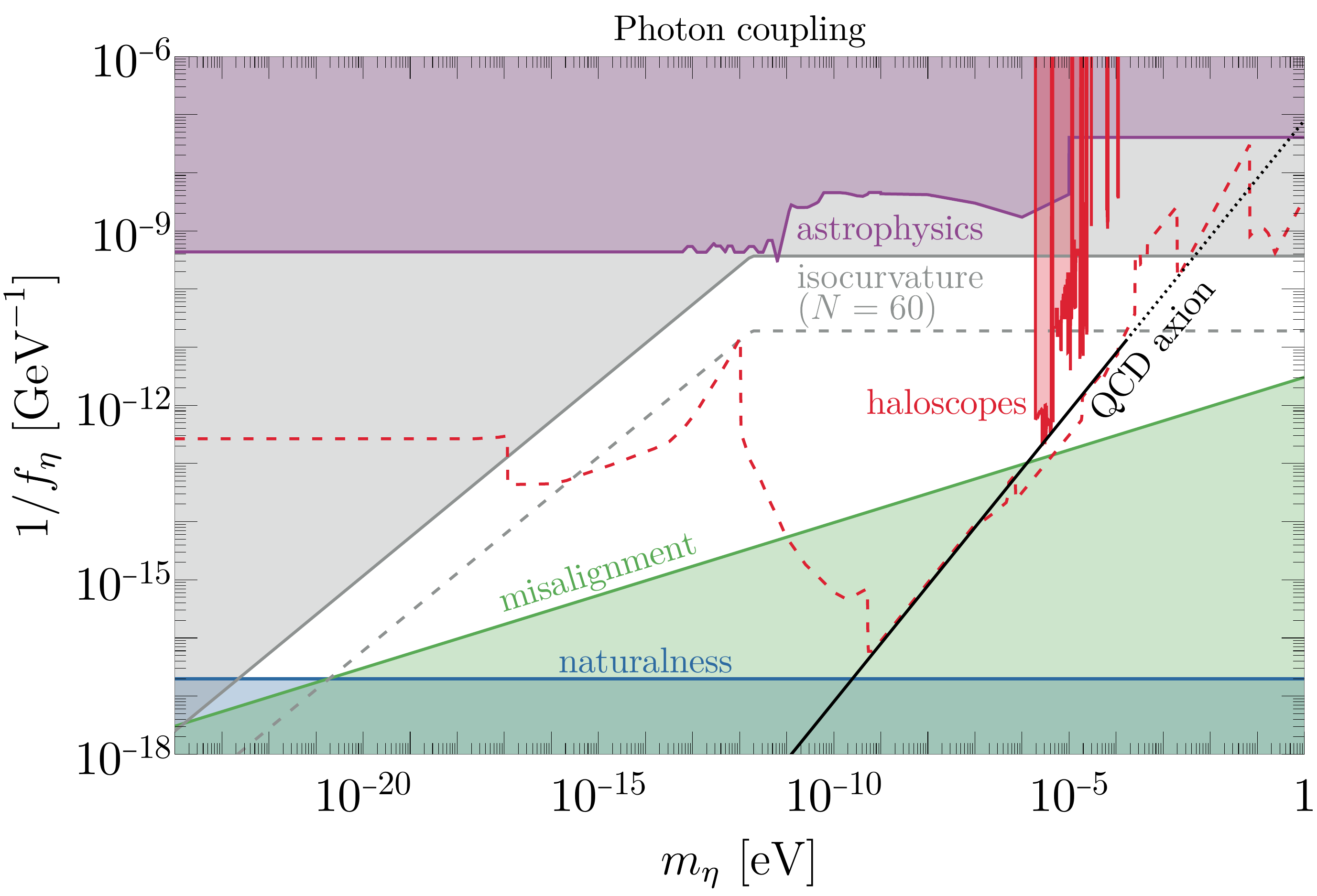}\quad
	\includegraphics[width=0.48\linewidth]{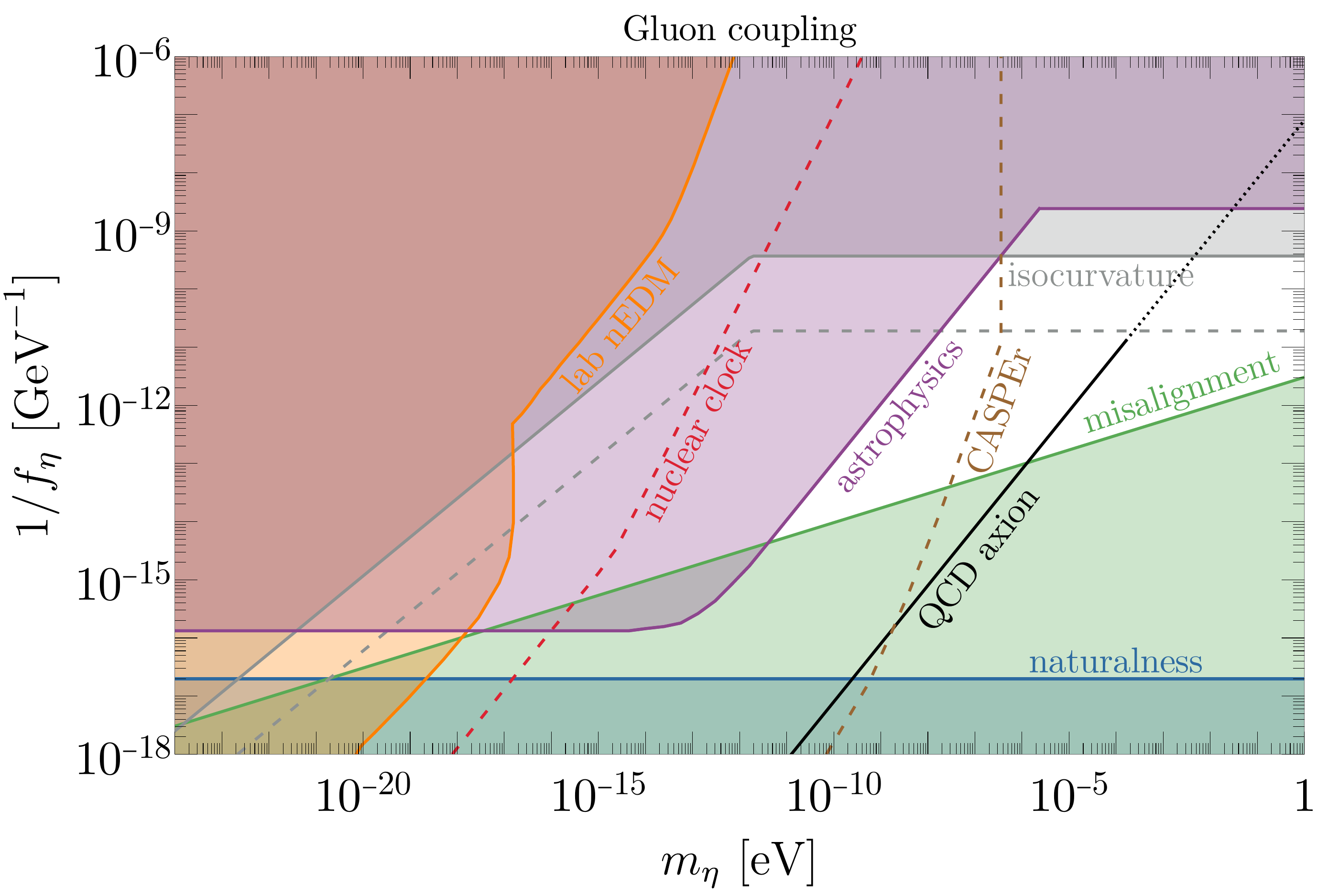}
        \caption{\label{fig:photongluon coupling}Constraints on the axion couplings to the photon (left) and the gluon (right). The couplings are given in Eq.~\eqref{eq:couplings}. The QCD axion falls along the black line; we dot the line for $m \gtrsim 10^{-4}$~eV, the maximum mass our mechanism can achieve for $N = 60$ e-folds of inflation (see Fig.~\ref{fig:parameter}). The gray region is excluded by bounds on isocurvature perturbations for $N = 60$ e-folds,  but this bound is relaxed for more e-folds. We also show a projection~\cite{Feix:2019lpo} for future isocurvature bounds from CMB-S4~\cite{PRISM:2013fvg,CMB-S4:2016ple} and SKA2~\cite{Cosmology-SWG:2015tjb,Weltman:2018zrl} with the dashed gray line. The green line corresponds to the misalignment production of ALP DM, while the blue region is excluded by the naturalness bound in Eq.~\eqref{eq:naturalness}. In the left panel we present astrophysical bounds on photon-axion conversion~\cite{Reynes:2021bpe,Reynolds:2019uqt,Marsh:2017yvc,Dessert:2020lil,Hoof:2022xbe,Fermi-LAT:2016nkz,Dessert:2022yqq,Noordhuis:2022ljw,Ayala:2014pea,Balkin:2020dsr} in purple, bound from axion haloscopes~\cite{ADMX:2009iij,ADMX:2018gho,ADMX:2019uok,ADMX:2021nhd,DePanfilis:1987dk,Hagmann:1990tj,HAYSTAC:2018rwy,HAYSTAC:2020kwv,HAYSTAC:2023cam,TASEH:2022vvu,Quiskamp:2022pks,Quiskamp:2023ehr,Lee:2020cfj,Jeong:2020cwz,CAPP:2020utb,Lee:2022mnc,Kim:2022hmg,Yi:2022fmn,Yang:2023yry,Kim:2023vpo,Adair:2022rtw} in red, and projections for future haloscopes~\cite{AxionLimits} as a dashed red line. In the right panel we show astrophysical bounds from supernova 1987a~\cite{Raffelt:2006cw,Caputo:2024oqc} and white dwarf cooling~\cite{Hook:2017psm,Balkin:2022qer} in purple, as well as bounds from oscillating neutron EDM experiments~\cite{Abel:2017rtm,Schulthess:2022pbp} in orange. We present projections for a $^{229}{\rm Th}$ nuclear clock~\cite{Kim:2022ype} and CASPEr-electric~\cite{Budker:2013hfa,JacksonKimball:2017elr} as dashed red and brown lines, respectively. All limits are adapted from~\cite{AxionLimits,Kim:2022ype}. }  
\end{figure*}

\subsection{QCD axion}
Things are different for the QCD axion, where $m_\eta$ is induced by nonperturbative QCD effects at the characteristic scale $\Lambda_{\rm QCD}\approx 200~{\rm MeV}$. Therefore, the QCD axion is strictly massless at the end of inflation (assuming $T_{\rm reh}>\Lambda_{\rm  QCD}$), and becomes nonrelativistic when $T\lesssim \Lambda_{\rm QCD}$. The present-day energy density can be estimated by $\langle \rho_\eta(\tau_0) \rangle= \langle \rho_\eta(\tau_{\rm e}) \rangle\left(a_{\rm e}/a_{\rm QCD}\right)^4\left(a_{\rm QCD}/{a_{0}}\right)^3$, where $a_{\rm QCD}$ is the scale factor at QCD phase transition. Therefore, we find the relic abundance of the QCD axion to be
\begin{align}
	\label{eq:abundance-QCD}
\Omega_\eta = \frac{g_{*0}}{4320\pi}\frac{\Lambda_{\rm  QCD}T_0^3 H_{\rm inf}^2}{M_{\rm Pl}^4 H_0^2}{\cal F}(\kappa)\,e^{N\left(2\nu-3\right)}\;,
\end{align}
which does not depend on the axion mass either. This should be understood as a crude estimate because we have neglected the temperature dependence of axion mass. If the temperature dependence is included, then axion can become nonrelativistic before the temperature drops below $\Lambda_{\rm  QCD}$.

Note that for the QCD axion, the breaking scale $f_\eta$ is related to the mass through $f_\eta m_\eta \approx \Lambda_{\rm QCD}^2$. Requiring that $f_\eta > H_{\rm inf}/2\pi$ (so that PQ symmetry is broken during inflation) then puts an upper bound on the QCD axion mass, which we present as black curves in Fig.~\ref{fig:parameter}. Combined with the backreaction constraint this implies a maximum QCD axion mass of order $10^{-4}$~eV for $N=60$ e-folds, although this can be relaxed by increasing the number of e-folds of inflation. Compared with the minimal misalignment mechanism without fine-tuning, where the QCD axion mass should be of order $10^{-6}~{\rm eV}$  to comprise all of the DM, our mechanism allows for a heavier QCD axion, corresponding to a larger coupling to SM particles. Lastly, in Fig.~\ref{fig:parameter} we also show the lower bound on the QCD axion mass from the requirement that $f_\eta < M_{\rm Pl}$.

\section{Phenomenology}
In Fig.~\ref{fig:photongluon coupling} we present constraints on our parameter space, focusing on the axion couplings to the photon and the gluon. For ALPs, we assume the couplings are
\begin{equation}\label{eq:couplings}
    \frac{\alpha_{\rm EM}}{8 \pi f_\eta} \eta F_{\mu\nu} \tilde{F}^{\mu\nu}, \quad \frac{\alpha_s}{8 \pi f_\eta} \eta G_{\mu\nu}^a \tilde{G}^{\mu\nu}_a
\end{equation}
up to an order-one factor which we neglect. Here $F_{\mu\nu}$ and $G_{\mu\nu}^a$ are the photon and gluon field strengths, respectively; $\alpha_{\rm EM}$ and $\alpha_s$ are the electromagnetic and strong fine-structure constants, respectively. For the QCD axion, the gluon coupling is given by Eq.~\eqref{eq:couplings}, but the photon coupling includes the usual model-dependent constant $E/N - 1.92$, where $E/N$ is the ratio of the electromagnetic and color anomalies of the axial current~\cite{ParticleDataGroup:2022pth}. For concreteness we set this factor to one for our plots, but one can easily rescale for an arbitrary value of $E/N$. We depict the parameter space occupied by the QCD axion as a black line in Fig.~\ref{fig:photongluon coupling}. For $N = 60$ e-folds of inflation the QCD axion cannot be heavier than about $10^{-4}$~eV, indicated by the dotting in Fig.~\ref{fig:photongluon coupling}, but heavier masses can be achieved with more e-folds (see Fig.~\ref{fig:parameter}).

Since we are considering a broken PQ scenario, isocurvature perturbations from the axion are important~\cite{Fox:2004kb,Beltran:2006sq,Acharya:2010zx,Marsh:2014qoa,Feix:2019lpo,Chen:2023txq}. We present the calculational details of the isocurvature bound in Appendix~\ref{app:isocurvature}. In Fig.~\ref{fig:photongluon coupling} we show this bound assuming $N = 60$ e-folds and an ${\cal O}(1)$ misalignment angle. Next-generation CMB surveys and $21$~cm experiments will improve sensitivity to isocurvature perturbations; we include an optimistic projection, adapted from~\cite{Feix:2019lpo}, combining CMB-S4~\cite{PRISM:2013fvg,CMB-S4:2016ple} and SKA2~\cite{Cosmology-SWG:2015tjb,Weltman:2018zrl}. We also remark that the isocurvature bound can be relaxed by increasing the number of e-folds.

There is also a naturalness bound on the decay constant $f_\eta$. The inflaton mass $m_\phi$ receives loop corrections from the axion coupling of order $f_\eta^4/(16\pi^2 m_\phi^2)$.\footnote{We are assuming that the only scale in the inflaton potential is $m_\phi$. One can repeat the analysis if the inflaton is a pNGB and the conclusion is unchanged.} Thus we require for naturalness $m_\phi^2 \gtrsim f_\eta^2/4\pi$, and since the inflaton energy density is of order $m_\phi^4$, this leads to a bound
\begin{equation}\label{eq:naturalness}
    \frac{f_\eta^4}{16\pi^2} \lesssim 3 H_{\rm inf}^2 M_{\rm Pl}^2 .
\end{equation}
Together with the upper bound on the inflationary Hubble scale ($H_{\rm inf} < 4.8 \times 10^{13}$~GeV), this gives an upper bound $f_\eta \lesssim 5 \times 10^{16}$~GeV, which we plot as a blue line in Fig.~\ref{fig:photongluon coupling}. One could probably evade this bound with appropriate inflationary model-building.

There are a variety of astrophysical and cosmological bounds on the DM mass that exclude DM lighter than about $10^{-22}$--$10^{-20}$~eV, including constraints from the Lyman $\alpha$-forest~\cite{Rogers:2020ltq}, pulsar timing array~\cite{NANOGrav:2023hvm}, star cluster observations~\cite{Marsh:2018zyw}, the Milky Way mass function~\cite{DES:2020fxi}, and black hole superradiance~\cite{Arvanitaki:2010sy,Mehta:2020kwu} (for a review see e.g.~\cite{Antypas:2022asj}). For clarity we have not presented these bounds in Fig.~\ref{fig:photongluon coupling}.

It is important to check that we do not have significant DM production from misalignment in addition to our mechanism. Recall that the usual misalignment mechanism predicts a relic abundance
\begin{align}
	\label{eq:misalignment}
\left(\frac{\Omega_\eta h^2}{0.12}\right)_{\rm mis.}\sim \left(\frac{m_\eta}{10^{-10}~{\rm eV}}\right)^{1/2}\left(\frac{f_\eta}{10^{14}~{\rm GeV}}\right)^2\;,
\end{align}
assuming an $\mathcal{O}(1)$ misalignment angle, where $h$ is the reduced Hubble constant. The green line in Fig.~\ref{fig:photongluon coupling} corresponds to where misalignment yields $\Omega_\eta h^2 \sim 0.12$. Unless one fine-tunes the misalignment angle to suppress this contribution to the relic density, our mechanism is only viable above this line. This is different than the usual misalignment story, where the ALP must lie on this line (or below, with fine-tuning).

Our mechanism allows for a lower PQ breaking scale than misalignment, which means the couplings to SM particles can be larger and easier to detect experimentally. In Fig.~\ref{fig:photongluon coupling} we present existing bounds on the axion-photon coupling from astrophysical limits on photon-axion conversion~\cite{Reynes:2021bpe,Reynolds:2019uqt,Marsh:2017yvc,Dessert:2020lil,Hoof:2022xbe,Fermi-LAT:2016nkz,Dessert:2022yqq,Noordhuis:2022ljw,Ayala:2014pea,Balkin:2020dsr} and from axion haloscopes~\cite{ADMX:2009iij,ADMX:2018gho,ADMX:2019uok,ADMX:2021nhd,DePanfilis:1987dk,Hagmann:1990tj,HAYSTAC:2018rwy,HAYSTAC:2020kwv,HAYSTAC:2023cam,TASEH:2022vvu,Quiskamp:2022pks,Quiskamp:2023ehr,Lee:2020cfj,Jeong:2020cwz,CAPP:2020utb,Lee:2022mnc,Kim:2022hmg,Yi:2022fmn,Yang:2023yry,Kim:2023vpo,Adair:2022rtw}, adapted from~\cite{AxionLimits}.
There is a wide swath of viable parameter space stretching across a large range of masses, as shown by the white regions in Fig.~\ref{fig:photongluon coupling}. We include projections for future axion haloscopes (adapted from~\cite{AxionLimits}), such as DANCE~\cite{Michimura:2019qxr}, SRF~\cite{Berlin:2020vrk}, DM-Radio~\cite{DMRadio:2022pkf}, twisted anyon cavity~\cite{Bourhill:2022alm} etc., which will effectively probe the photon coupling parameter space. For the axion-gluon coupling we show astrophysical bounds~\cite{Raffelt:2006cw,Caputo:2024oqc,Hook:2017psm,Balkin:2022qer} and constraints from neutron EDM oscillation experiments~\cite{Abel:2017rtm,Schulthess:2022pbp}, adapted from~\cite{AxionLimits,Kim:2022ype}. The gluon coupling is heavily constrained for masses smaller than $10^{-11}$~eV. Of course, our framework does not require this coupling to be present for an ultralight ALP, as the coefficient of the $\eta G \tilde{G}$ operator is a free parameter of the theory (the generic coupling induced by gravity is Planck-suppressed). The existing constraints simply mean that this type of coupling is disfavored; one could instead consider an ALP that only couples to the photon at tree-level, for instance. We also show projections for a future $^{229}{\rm Th}$ nuclear clock~\cite{Kim:2022ype} and for CASPEr-electric~\cite{Budker:2013hfa,JacksonKimball:2017elr} in Fig.~\ref{fig:photongluon coupling}, which can further probe/constrain the gluon coupling parameter space.

\section{Conclusions}
In this work, we proposed a new mechanism to produce axion DM through a quantum phase transition that is driven by inflation. The kinetic coupling to the inflaton breaks the scale-invariant power spectrum of a massless axion, and leads to a phase transition from the conformal symmetric phase to the broken phase. We find that the phase transition is completely controlled by the shape of the coupling through the effective curvature $\kappa$ defined in Eq.~(\ref{eq:curvaturedef}), which plays the role of an order parameter. If $\kappa$ is positive, the axion spectrum is broken to a red tilt and the axion abundance gets an exponential enhancement during inflation, compensating the suppression from the small axion mass. The viable parameter space for axion DM in terms of $\kappa$ is shown in Fig.~\ref{fig:parameter}, which is a general result independent of the ultraviolet (UV) origin of the kinetic coupling.

Our mechanism allows for a much lower axion decay constant than misalignment, and covers a large range of scales for ultralight DM, from sub-eV down to $10^{-24}~{\rm eV}$ (for $N = 60$ e-folds, or even lower with more e-folds). The viable parameter space can be probed by future axion experiments, including haloscopes, nuclear clocks, CASPEr, and next-generation CMB surveys as well as $21$~cm telescopes, as shown in Fig.~\ref{fig:photongluon coupling}. Our mechanism is applicable to both the QCD axion and ALPs. In addition, we expect that it can also be applied to other bosonic ultralight DM scenarios.
A more detailed investigation of the phenomenology of our mechanism will be carried out in separate work~\cite{Ismail:2024}.

\begin{acknowledgments}
We would like to thank Haipeng An, Abhishek Banerjee, Ying-Ying Li, Yuichiro Nakai, Maxim Perelstein, Gilad Perez, Maximilian Ruhdorfer, Konstantin Springmann, Stefan Stelzl, Xin Wang, Ziwei Wang, and Taewook Youn for helpful discussions and feedback. We thank Les Houches workshop PhysTeV 2023 for inspiring us to start this work. AI is supported in part by the NSF grant PHY-2014071 and in part by NSERC, funding reference number 557763. SJL and BY are supported by the Samsung Science Technology Foundation under Project Number SSTF-BA2201-06. SJL is also supported by the National Research Foundation of Korea (NRF) funded by the Korea government (MEST).
\end{acknowledgments}

\appendix
\section{Realizations of kinetic coupling}
\label{app:examples}
During the general discussion in the main text, we did not specify the origin of kinetic coupling. For any UV model with a given kinetic function $K(\phi)$ and inflaton potential $V(\phi)$, one could directly compute the effective curvature $\kappa$ using Eq.~(\ref{eq:curvature}) and map it onto the general result shown in Fig.~\ref{fig:parameter}. 
In this appendix, we discuss some concrete examples of the kinetic coupling, and derive the conditions to realize a positive effective curvature, which can lead to an exponential enhancement of the axion abundance  during inflation.

\subsection{Effective operator}
\label{appsub:EFT}
As a first example, we assume the kinetic coupling comes from some higher-order effective operator\footnote{For other high-ordered operators, including dimension-five operator, we have similar results.}:
\begin{align}
	K(\phi) = 1 + \frac{C_6}{M_{\rm Pl}^2}\left(\phi^2-\phi_{\rm e}^2\right)\;,
\end{align}
where $\phi_{\rm e}$ is the field value at the end of inflation and $C_6$ is the dimensionless Wilson coefficient determined by unknown UV physics. In addition, $C_6$ should satisfy $\left|C_6 \phi^2\right|<M_{\rm Pl}^2$ during inflation to ensure the validity of the effective field theory. Taking the quadratic inflaton potential $V(\phi)=m_\phi^2 \phi^2/2$ as a toy model, using Eq.~(\ref{eq:curvature}), we obtain
\begin{align}
	\kappa\approx -4C_6\left(3-\epsilon\right)\approx -12C_6\;.
\end{align}
It is interesting to notice that although the kinetic coupling from the higher-order operator is suppressed by $1/M_{\rm Pl}^{2}$, the effective curvature $\kappa$, which breaks scale-invariant spectrum and drives phase transition, is not suppressed. 
Therefore, to have an exponential enhancement, $C_6$ should be negative during inflation.

\subsection{Power law}
\label{appsub:powerlaw}
Next we consider the scenario where kinetic function is proportional to some power of the scale factor:
\begin{align}
	\label{eq:powerlaw}
K(\phi) = \left(a/a_{\rm e}\right)^n\;,
\end{align}
with $n$ an arbitrary real number. This simple and compact form has been widely used in the literature to study the kinetic coupling between inflaton and dark photon~\cite{Martin:2007ue,Watanabe:2009ct,Namba:2012gg,Bartolo:2012sd,Nakayama:2019rhg,Nakayama:2020rka,Nakai:2020cfw}, which can be used to, e.g., generate large-scale magnetic fields in the universe~\cite{Martin:2007ue}, or produce dark photon as ultralight DM~\cite{Nakai:2020cfw}. Eq.~(\ref{eq:powerlaw}) can be realized, under the slow-roll approximation, through
\begin{align}
	\label{eq:powerlaw2}
K(\phi)=\exp\left[-\frac{n}{M_{\rm Pl}^2}\int_{\phi_{\rm e}}^{\phi}{\rm d}\phi'\frac{V\left(\phi'\right)}{V_\phi\left(\phi'\right)}\right]\;,
\end{align}
with an arbitrary inflaton potential $V(\phi)$. From Eq.~(\ref{eq:powerlaw2}), one can compute the effective curvature using Eq.~(\ref{eq:curvature}):
\begin{align}
\kappa \approx n\left(n+3-2\epsilon+\eta_V\right)\approx n\left(n+3\right)\;,
\end{align}
where $\eta_V\equiv M_{\rm Pl}^2 V_{\phi\phi}/V \ll 1$ is the second slow-roll parameter. Therefore, in order to realize a positive effective curvature, the power index $n$ should satisfy $n>0$ or $n<-3$.

\subsection{Radial mode as inflaton}
\label{appsub:radial}
Finally, let us consider an interesting model where the radial mode of the PQ scalar plays a role of the inflaton~\cite{Fairbairn:2014zta, Lee:2023dtw}. The key observation is that the kinetic term of PQ scalar $\chi=\rho\,e^{{\rm i}\eta/f_\eta}/\sqrt{2}$ naturally leads to a coupling between inflaton $\rho$ and axion kinetic term:
\begin{align}
	\left|\partial_\mu \chi\right|^2=\frac{1}{2}\left[\left(\partial_\mu \rho\right)^2+\frac{\rho^2}{f_\eta^2}\left(\partial_\mu \eta\right)^2\right]\;.
\end{align}
During inflation, we have $\rho\gg f_\eta$ and the axion kinetic coupling is significant. As $\rho$ rolls down along the potential and tends to the vacuum expectation value $f_\eta$, the inflation ends and the axion kinetic term reduces to the canonical form. The great advantage of this model is that the isocurvature fluctuation produced from axion is highly suppressed. This is because the effective PQ breaking scale during inflation is $f_{\rm eff}=\rho$, therefore the isocurvature fluctuation is suppressed by $f_\eta^2/\rho^2\ll 1$~\cite{Fairbairn:2014zta}.

The action in this model reads
\begin{align}
	{\cal S}=\int {\rm d}^4 x&\sqrt{-g}\left[\frac{M_{\rm Pl}^2}{2}R\left(1+\xi\frac{\rho^2}{M_{\rm Pl}^2}\right)-\frac{1}{2}g^{\mu\nu}\partial_\mu \rho \partial_\nu \rho\right.\nonumber\\
	&\left.-\frac{1}{2}\frac{\rho^2}{f_\eta^2}g^{\mu\nu}\partial_\mu\eta\partial_\nu\eta-\frac{\lambda}{4}\left(\rho^2-f_\eta^2\right)^2\right]\;,
\end{align}
where a non-minimal coupling term of the inflaton to gravity is added in order to be compatible with the CMB constraints on the tensor-to-scalar ratio~\cite{Fairbairn:2014zta,Linde:2011nh}. After transforming from Jordan frame to Einstein frame (which has a canonical gravity term), the kinetic function and the inflaton potential are given by
\begin{align}
	K(\rho) &= \frac{\rho}{f_\eta\sqrt{\left(1+\xi \rho^2/M_{\rm Pl}^2\right)}}\;,\\
	V(\rho) & = 
	\frac{\lambda\left(\rho^2-f_\eta^2\right)^2}{4\left(1+\xi \rho^2/M_{\rm Pl}^2\right)^2}\;.
\end{align}
Using Eq.~(\ref{eq:curvature}), one can compute the effective curvature $\kappa$:
\begin{align}
	\label{eq:kappa-radial}
\kappa \approx& -4 q^4 \left[3 \xi ^2 (6 \xi +1)^2+\left(24 \xi ^2+8 \xi +3\right) q^4\right.\nonumber\\
&\left. +2 \xi  \left(24 \xi ^2+22 \xi +3\right) q^2\right]/\left(6 \xi^2+\xi +q^2\right)^3\;,
\end{align}
where $q\equiv M_{\rm Pl}/\rho$. To have a positive $\kappa$ during inflation, we find that the non-minimal coupling parameter $\xi$ should satisfy $-1/6 < \xi < 0 $. Note that though the effective curvature in Eq.~(\ref{eq:kappa-radial}) depends on the inflaton, under the slow-roll approximation, its change during inflation is of ${\cal O}(1)$, which is much slower than the exponential growth of the axion field. Consequently, it is self-consistent for us to treat $\kappa$ as a constant in Eq.~(\ref{eq:EOM-k}).

\section{Isocurvature bound}
\label{app:isocurvature}
Here we provide calculational details of the isocurvature bound~\cite{Fox:2004kb,Beltran:2006sq,Acharya:2010zx,Marsh:2014qoa,Feix:2019lpo,Chen:2023txq}. Recall that the isocurvature (entropy) mode measures the deviation from the adiabatic mode of single-field inflation, which can be parametrized as
\begin{align}
	\beta_{\rm iso} \equiv A_{\rm iso}/\left(A_{\rm s}+A_{\rm iso}\right) \approx A_{\rm iso}/A_{\rm s}\;,
\end{align} 
where $A_{\rm s}$ is the scalar amplitude from the adiabatic mode and $A_{\rm iso}$ is the isocurvature perturbation caused by the axion. Since the axion spectrum during inflation is dominated by the superhorizon mode, we only need to consider the isocurvature perturbation caused by $k_{\rm min}$. Unlike the scale-invariant spectrum for a canonical massless scalar, the kinetic coupling enhances the axion fluctuation as
\begin{align}
	\left\langle \delta \eta^2 \right\rangle =\left(H_{\rm inf}/2\pi\right)^2\left(k_*/k_{\rm min}\right)^{2\nu-3}\;,
\end{align}
where $k_*= 0.05~{\rm Mpc}^{-1}$ is the pivot scale. So the axion isocurvature perturbation is given by (assuming it constitutes the entire cold DM)
\begin{align}
	\beta_{\rm iso} = \frac{1}{A_{\rm s}}\frac{H_{\rm inf}^2}{\pi^2 \eta_{\rm i}^2}\left(\frac{k_*}{k_{\rm min}}\right)^{2\nu-3}\;,
\end{align}
with $\eta_{\rm i}^{}=f_\eta^{} \theta_{\rm i}^{}$ the initial field displacement and $\theta_{\rm i}$ the initial misalignment angle. The latest Planck measurements give $\beta_{\rm iso}<0.038$~\cite{Planck:2018vyg}, which puts a lower bound on the ratio between PQ breaking scale and inflationary Hubble scale:
\begin{align}
	\label{eq:isocurvature}
	f_\eta \theta_{\rm i}/H_{\rm inf} > 3.5 \times 10^{4}/\left(\sqrt{\kappa}H_0/k_*\right)^{\nu-3/2} \;,
\end{align}
where we have taken the initial horizon to be the horizon today to fix the upper bound of the minimum mode, i.e., $k_{\rm min}=\sqrt{\kappa} a_{\rm i}H_{\rm inf}\leq \sqrt{\kappa} a_0 H_0$. Thanks to the back-reaction constraint, the effective curvature $\kappa$ cannot be too large. Hence the modification from the kinetic coupling in Eq.~(\ref{eq:isocurvature}) is of ${\cal O}(1)$. More specifically, for $0<\kappa<1$, we have $0.19<\left(\sqrt{\kappa}H_0/k_*\right)^{\nu-3/2}<1$. 

The isocurvature bound in Eq.~\eqref{eq:isocurvature} is shown in Fig.~\ref{fig:photongluon coupling} as gray lines, taking an ${\cal O}(1)$ initial misalignment angle and $N=60$. Note that this bound is easily relaxed for a larger number of e-folds, which allows a lower Hubble during inflation. Future experiments like CMB-S4~\cite{PRISM:2013fvg,CMB-S4:2016ple} and SKA2~\cite{Cosmology-SWG:2015tjb,Weltman:2018zrl} will improve sensitivity to CMB spectral distortions and thus could strengthen the isocurvature bound. We include a projection adapted from~\cite{Feix:2019lpo} in Fig.~\ref{fig:photongluon coupling}.

%

\bibliography{ref}

\begin{thebibliography}{114}%
\makeatletter
\providecommand \@ifxundefined [1]{%
 \@ifx{#1\undefined}
}%
\providecommand \@ifnum [1]{%
 \ifnum #1\expandafter \@firstoftwo
 \else \expandafter \@secondoftwo
 \fi
}%
\providecommand \@ifx [1]{%
 \ifx #1\expandafter \@firstoftwo
 \else \expandafter \@secondoftwo
 \fi
}%
\providecommand \natexlab [1]{#1}%
\providecommand \enquote  [1]{``#1''}%
\providecommand \bibnamefont  [1]{#1}%
\providecommand \bibfnamefont [1]{#1}%
\providecommand \citenamefont [1]{#1}%
\providecommand \href@noop [0]{\@secondoftwo}%
\providecommand \href [0]{\begingroup \@sanitize@url \@href}%
\providecommand \@href[1]{\@@startlink{#1}\@@href}%
\providecommand \@@href[1]{\endgroup#1\@@endlink}%
\providecommand \@sanitize@url [0]{\catcode `\\12\catcode `\$12\catcode
  `\&12\catcode `\#12\catcode `\^12\catcode `\_12\catcode `\%12\relax}%
\providecommand \@@startlink[1]{}%
\providecommand \@@endlink[0]{}%
\providecommand \url  [0]{\begingroup\@sanitize@url \@url }%
\providecommand \@url [1]{\endgroup\@href {#1}{\urlprefix }}%
\providecommand \urlprefix  [0]{URL }%
\providecommand \Eprint [0]{\href }%
\providecommand \doibase [0]{http://dx.doi.org/}%
\providecommand \selectlanguage [0]{\@gobble}%
\providecommand \bibinfo  [0]{\@secondoftwo}%
\providecommand \bibfield  [0]{\@secondoftwo}%
\providecommand \translation [1]{[#1]}%
\providecommand \BibitemOpen [0]{}%
\providecommand \bibitemStop [0]{}%
\providecommand \bibitemNoStop [0]{.\EOS\space}%
\providecommand \EOS [0]{\spacefactor3000\relax}%
\providecommand \BibitemShut  [1]{\csname bibitem#1\endcsname}%
\let\auto@bib@innerbib\@empty
\bibitem [{\citenamefont {Ferreira}(2021)}]{Ferreira:2020fam}%
  \BibitemOpen
  \bibfield  {author} {\bibinfo {author} {\bibfnamefont {E.~G.~M.}\
  \bibnamefont {Ferreira}},\ }\href {\doibase 10.1007/s00159-021-00135-6}
  {\bibfield  {journal} {\bibinfo  {journal} {Astron. Astrophys. Rev.}\
  }\textbf {\bibinfo {volume} {29}},\ \bibinfo {pages} {7} (\bibinfo {year}
  {2021})},\ \Eprint {http://arxiv.org/abs/2005.03254} {arXiv:2005.03254
  [astro-ph.CO]} \BibitemShut {NoStop}%
\bibitem [{\citenamefont {Hui}(2021)}]{Hui:2021tkt}%
  \BibitemOpen
  \bibfield  {author} {\bibinfo {author} {\bibfnamefont {L.}~\bibnamefont
  {Hui}},\ }\href {\doibase 10.1146/annurev-astro-120920-010024} {\bibfield
  {journal} {\bibinfo  {journal} {Ann. Rev. Astron. Astrophys.}\ }\textbf
  {\bibinfo {volume} {59}},\ \bibinfo {pages} {247} (\bibinfo {year} {2021})},\
  \Eprint {http://arxiv.org/abs/2101.11735} {arXiv:2101.11735 [astro-ph.CO]}
  \BibitemShut {NoStop}%
\bibitem [{\citenamefont {Antypas}\ \emph {et~al.}(2022)\citenamefont {Antypas}
  \emph {et~al.}}]{Antypas:2022asj}%
  \BibitemOpen
  \bibfield  {author} {\bibinfo {author} {\bibfnamefont {D.}~\bibnamefont
  {Antypas}} \emph {et~al.},\ }\href@noop {} {\  (\bibinfo {year} {2022})},\
  \Eprint {http://arxiv.org/abs/2203.14915} {arXiv:2203.14915 [hep-ex]}
  \BibitemShut {NoStop}%
\bibitem [{\citenamefont {Kim}\ and\ \citenamefont
  {Mitridate}(2023)}]{Kim:2023kyy}%
  \BibitemOpen
  \bibfield  {author} {\bibinfo {author} {\bibfnamefont {H.}~\bibnamefont
  {Kim}}\ and\ \bibinfo {author} {\bibfnamefont {A.}~\bibnamefont
  {Mitridate}},\ }\href@noop {} {\  (\bibinfo {year} {2023})},\ \Eprint
  {http://arxiv.org/abs/2312.12225} {arXiv:2312.12225 [hep-ph]} \BibitemShut
  {NoStop}%
\bibitem [{\citenamefont {Kim}\ and\ \citenamefont
  {Perez}(2024)}]{Kim:2022ype}%
  \BibitemOpen
  \bibfield  {author} {\bibinfo {author} {\bibfnamefont {H.}~\bibnamefont
  {Kim}}\ and\ \bibinfo {author} {\bibfnamefont {G.}~\bibnamefont {Perez}},\
  }\href {\doibase 10.1103/PhysRevD.109.015005} {\bibfield  {journal} {\bibinfo
   {journal} {Phys. Rev. D}\ }\textbf {\bibinfo {volume} {109}},\ \bibinfo
  {pages} {015005} (\bibinfo {year} {2024})},\ \Eprint
  {http://arxiv.org/abs/2205.12988} {arXiv:2205.12988 [hep-ph]} \BibitemShut
  {NoStop}%
\bibitem [{\citenamefont {Flambaum}\ and\ \citenamefont
  {Samsonov}(2023)}]{Flambaum:2023bnw}%
  \BibitemOpen
  \bibfield  {author} {\bibinfo {author} {\bibfnamefont {V.~V.}\ \bibnamefont
  {Flambaum}}\ and\ \bibinfo {author} {\bibfnamefont {I.~B.}\ \bibnamefont
  {Samsonov}},\ }\href {\doibase 10.1103/PhysRevD.108.075022} {\bibfield
  {journal} {\bibinfo  {journal} {Phys. Rev. D}\ }\textbf {\bibinfo {volume}
  {108}},\ \bibinfo {pages} {075022} (\bibinfo {year} {2023})},\ \Eprint
  {http://arxiv.org/abs/2302.11167} {arXiv:2302.11167 [hep-ph]} \BibitemShut
  {NoStop}%
\bibitem [{\citenamefont {Flambaum}\ and\ \citenamefont
  {Mansour}(2023)}]{Flambaum:2023drb}%
  \BibitemOpen
  \bibfield  {author} {\bibinfo {author} {\bibfnamefont {V.~V.}\ \bibnamefont
  {Flambaum}}\ and\ \bibinfo {author} {\bibfnamefont {A.~J.}\ \bibnamefont
  {Mansour}},\ }\href {\doibase 10.1103/PhysRevLett.131.113004} {\bibfield
  {journal} {\bibinfo  {journal} {Phys. Rev. Lett.}\ }\textbf {\bibinfo
  {volume} {131}},\ \bibinfo {pages} {113004} (\bibinfo {year} {2023})},\
  \Eprint {http://arxiv.org/abs/2304.04469} {arXiv:2304.04469 [hep-ph]}
  \BibitemShut {NoStop}%
\bibitem [{\citenamefont {Zhao}\ \emph {et~al.}(2024)\citenamefont {Zhao},
  \citenamefont {Liu},\ and\ \citenamefont {Mei}}]{Zhao:2024tvg}%
  \BibitemOpen
  \bibfield  {author} {\bibinfo {author} {\bibfnamefont {W.}~\bibnamefont
  {Zhao}}, \bibinfo {author} {\bibfnamefont {H.}~\bibnamefont {Liu}}, \ and\
  \bibinfo {author} {\bibfnamefont {X.}~\bibnamefont {Mei}},\ }\href@noop {} {\
   (\bibinfo {year} {2024})},\ \Eprint {http://arxiv.org/abs/2401.17055}
  {arXiv:2401.17055 [hep-ph]} \BibitemShut {NoStop}%
\bibitem [{\citenamefont {Hu}\ \emph {et~al.}(2000)\citenamefont {Hu},
  \citenamefont {Barkana},\ and\ \citenamefont {Gruzinov}}]{Hu:2000ke}%
  \BibitemOpen
  \bibfield  {author} {\bibinfo {author} {\bibfnamefont {W.}~\bibnamefont
  {Hu}}, \bibinfo {author} {\bibfnamefont {R.}~\bibnamefont {Barkana}}, \ and\
  \bibinfo {author} {\bibfnamefont {A.}~\bibnamefont {Gruzinov}},\ }\href
  {\doibase 10.1103/PhysRevLett.85.1158} {\bibfield  {journal} {\bibinfo
  {journal} {Phys. Rev. Lett.}\ }\textbf {\bibinfo {volume} {85}},\ \bibinfo
  {pages} {1158} (\bibinfo {year} {2000})},\ \Eprint
  {http://arxiv.org/abs/astro-ph/0003365} {arXiv:astro-ph/0003365} \BibitemShut
  {NoStop}%
\bibitem [{\citenamefont {Hui}\ \emph {et~al.}(2017)\citenamefont {Hui},
  \citenamefont {Ostriker}, \citenamefont {Tremaine},\ and\ \citenamefont
  {Witten}}]{Hui:2016ltb}%
  \BibitemOpen
  \bibfield  {author} {\bibinfo {author} {\bibfnamefont {L.}~\bibnamefont
  {Hui}}, \bibinfo {author} {\bibfnamefont {J.~P.}\ \bibnamefont {Ostriker}},
  \bibinfo {author} {\bibfnamefont {S.}~\bibnamefont {Tremaine}}, \ and\
  \bibinfo {author} {\bibfnamefont {E.}~\bibnamefont {Witten}},\ }\href
  {\doibase 10.1103/PhysRevD.95.043541} {\bibfield  {journal} {\bibinfo
  {journal} {Phys. Rev. D}\ }\textbf {\bibinfo {volume} {95}},\ \bibinfo
  {pages} {043541} (\bibinfo {year} {2017})},\ \Eprint
  {http://arxiv.org/abs/1610.08297} {arXiv:1610.08297 [astro-ph.CO]}
  \BibitemShut {NoStop}%
\bibitem [{\citenamefont {Amruth}\ \emph {et~al.}(2023)\citenamefont {Amruth}
  \emph {et~al.}}]{Amruth:2023xqj}%
  \BibitemOpen
  \bibfield  {author} {\bibinfo {author} {\bibfnamefont {A.}~\bibnamefont
  {Amruth}} \emph {et~al.},\ }\href {\doibase 10.1038/s41550-023-01943-9}
  {\bibfield  {journal} {\bibinfo  {journal} {Nature Astron.}\ }\textbf
  {\bibinfo {volume} {7}},\ \bibinfo {pages} {736} (\bibinfo {year} {2023})},\
  \Eprint {http://arxiv.org/abs/2304.09895} {arXiv:2304.09895 [astro-ph.CO]}
  \BibitemShut {NoStop}%
\bibitem [{\citenamefont {Weinberg}(1978)}]{Weinberg:1977ma}%
  \BibitemOpen
  \bibfield  {author} {\bibinfo {author} {\bibfnamefont {S.}~\bibnamefont
  {Weinberg}},\ }\href {\doibase 10.1103/PhysRevLett.40.223} {\bibfield
  {journal} {\bibinfo  {journal} {Phys. Rev. Lett.}\ }\textbf {\bibinfo
  {volume} {40}},\ \bibinfo {pages} {223} (\bibinfo {year} {1978})}\BibitemShut
  {NoStop}%
\bibitem [{\citenamefont {Wilczek}(1978)}]{Wilczek:1977pj}%
  \BibitemOpen
  \bibfield  {author} {\bibinfo {author} {\bibfnamefont {F.}~\bibnamefont
  {Wilczek}},\ }\href {\doibase 10.1103/PhysRevLett.40.279} {\bibfield
  {journal} {\bibinfo  {journal} {Phys. Rev. Lett.}\ }\textbf {\bibinfo
  {volume} {40}},\ \bibinfo {pages} {279} (\bibinfo {year} {1978})}\BibitemShut
  {NoStop}%
\bibitem [{\citenamefont {Peccei}\ and\ \citenamefont
  {Quinn}(1977{\natexlab{a}})}]{Peccei:1977hh}%
  \BibitemOpen
  \bibfield  {author} {\bibinfo {author} {\bibfnamefont {R.~D.}\ \bibnamefont
  {Peccei}}\ and\ \bibinfo {author} {\bibfnamefont {H.~R.}\ \bibnamefont
  {Quinn}},\ }\href {\doibase 10.1103/PhysRevLett.38.1440} {\bibfield
  {journal} {\bibinfo  {journal} {Phys. Rev. Lett.}\ }\textbf {\bibinfo
  {volume} {38}},\ \bibinfo {pages} {1440} (\bibinfo {year}
  {1977}{\natexlab{a}})}\BibitemShut {NoStop}%
\bibitem [{\citenamefont {Peccei}\ and\ \citenamefont
  {Quinn}(1977{\natexlab{b}})}]{Peccei:1977ur}%
  \BibitemOpen
  \bibfield  {author} {\bibinfo {author} {\bibfnamefont {R.~D.}\ \bibnamefont
  {Peccei}}\ and\ \bibinfo {author} {\bibfnamefont {H.~R.}\ \bibnamefont
  {Quinn}},\ }\href {\doibase 10.1103/PhysRevD.16.1791} {\bibfield  {journal}
  {\bibinfo  {journal} {Phys. Rev. D}\ }\textbf {\bibinfo {volume} {16}},\
  \bibinfo {pages} {1791} (\bibinfo {year} {1977}{\natexlab{b}})}\BibitemShut
  {NoStop}%
\bibitem [{\citenamefont {Marsh}(2016)}]{Marsh:2015xka}%
  \BibitemOpen
  \bibfield  {author} {\bibinfo {author} {\bibfnamefont {D.~J.~E.}\
  \bibnamefont {Marsh}},\ }\href {\doibase 10.1016/j.physrep.2016.06.005}
  {\bibfield  {journal} {\bibinfo  {journal} {Phys. Rept.}\ }\textbf {\bibinfo
  {volume} {643}},\ \bibinfo {pages} {1} (\bibinfo {year} {2016})},\ \Eprint
  {http://arxiv.org/abs/1510.07633} {arXiv:1510.07633 [astro-ph.CO]}
  \BibitemShut {NoStop}%
\bibitem [{\citenamefont {Di~Luzio}\ \emph {et~al.}(2020)\citenamefont
  {Di~Luzio}, \citenamefont {Giannotti}, \citenamefont {Nardi},\ and\
  \citenamefont {Visinelli}}]{DiLuzio:2020wdo}%
  \BibitemOpen
  \bibfield  {author} {\bibinfo {author} {\bibfnamefont {L.}~\bibnamefont
  {Di~Luzio}}, \bibinfo {author} {\bibfnamefont {M.}~\bibnamefont {Giannotti}},
  \bibinfo {author} {\bibfnamefont {E.}~\bibnamefont {Nardi}}, \ and\ \bibinfo
  {author} {\bibfnamefont {L.}~\bibnamefont {Visinelli}},\ }\href {\doibase
  10.1016/j.physrep.2020.06.002} {\bibfield  {journal} {\bibinfo  {journal}
  {Phys. Rept.}\ }\textbf {\bibinfo {volume} {870}},\ \bibinfo {pages} {1}
  (\bibinfo {year} {2020})},\ \Eprint {http://arxiv.org/abs/2003.01100}
  {arXiv:2003.01100 [hep-ph]} \BibitemShut {NoStop}%
\bibitem [{\citenamefont {Dine}\ and\ \citenamefont
  {Fischler}(1983)}]{Dine:1982ah}%
  \BibitemOpen
  \bibfield  {author} {\bibinfo {author} {\bibfnamefont {M.}~\bibnamefont
  {Dine}}\ and\ \bibinfo {author} {\bibfnamefont {W.}~\bibnamefont
  {Fischler}},\ }\href {\doibase 10.1016/0370-2693(83)90639-1} {\bibfield
  {journal} {\bibinfo  {journal} {Phys. Lett. B}\ }\textbf {\bibinfo {volume}
  {120}},\ \bibinfo {pages} {137} (\bibinfo {year} {1983})}\BibitemShut
  {NoStop}%
\bibitem [{\citenamefont {Preskill}\ \emph {et~al.}(1983)\citenamefont
  {Preskill}, \citenamefont {Wise},\ and\ \citenamefont
  {Wilczek}}]{Preskill:1982cy}%
  \BibitemOpen
  \bibfield  {author} {\bibinfo {author} {\bibfnamefont {J.}~\bibnamefont
  {Preskill}}, \bibinfo {author} {\bibfnamefont {M.~B.}\ \bibnamefont {Wise}},
  \ and\ \bibinfo {author} {\bibfnamefont {F.}~\bibnamefont {Wilczek}},\ }\href
  {\doibase 10.1016/0370-2693(83)90637-8} {\bibfield  {journal} {\bibinfo
  {journal} {Phys. Lett. B}\ }\textbf {\bibinfo {volume} {120}},\ \bibinfo
  {pages} {127} (\bibinfo {year} {1983})}\BibitemShut {NoStop}%
\bibitem [{\citenamefont {Arias}\ \emph {et~al.}(2012)\citenamefont {Arias},
  \citenamefont {Cadamuro}, \citenamefont {Goodsell}, \citenamefont {Jaeckel},
  \citenamefont {Redondo},\ and\ \citenamefont {Ringwald}}]{Arias:2012az}%
  \BibitemOpen
  \bibfield  {author} {\bibinfo {author} {\bibfnamefont {P.}~\bibnamefont
  {Arias}}, \bibinfo {author} {\bibfnamefont {D.}~\bibnamefont {Cadamuro}},
  \bibinfo {author} {\bibfnamefont {M.}~\bibnamefont {Goodsell}}, \bibinfo
  {author} {\bibfnamefont {J.}~\bibnamefont {Jaeckel}}, \bibinfo {author}
  {\bibfnamefont {J.}~\bibnamefont {Redondo}}, \ and\ \bibinfo {author}
  {\bibfnamefont {A.}~\bibnamefont {Ringwald}},\ }\href {\doibase
  10.1088/1475-7516/2012/06/013} {\bibfield  {journal} {\bibinfo  {journal}
  {JCAP}\ }\textbf {\bibinfo {volume} {06}},\ \bibinfo {pages} {013} (\bibinfo
  {year} {2012})},\ \Eprint {http://arxiv.org/abs/1201.5902} {arXiv:1201.5902
  [hep-ph]} \BibitemShut {NoStop}%
\bibitem [{\citenamefont {Grilli~di Cortona}\ \emph {et~al.}(2016)\citenamefont
  {Grilli~di Cortona}, \citenamefont {Hardy}, \citenamefont {Pardo~Vega},\ and\
  \citenamefont {Villadoro}}]{GrillidiCortona:2015jxo}%
  \BibitemOpen
  \bibfield  {author} {\bibinfo {author} {\bibfnamefont {G.}~\bibnamefont
  {Grilli~di Cortona}}, \bibinfo {author} {\bibfnamefont {E.}~\bibnamefont
  {Hardy}}, \bibinfo {author} {\bibfnamefont {J.}~\bibnamefont {Pardo~Vega}}, \
  and\ \bibinfo {author} {\bibfnamefont {G.}~\bibnamefont {Villadoro}},\ }\href
  {\doibase 10.1007/JHEP01(2016)034} {\bibfield  {journal} {\bibinfo  {journal}
  {JHEP}\ }\textbf {\bibinfo {volume} {01}},\ \bibinfo {pages} {034} (\bibinfo
  {year} {2016})},\ \Eprint {http://arxiv.org/abs/1511.02867} {arXiv:1511.02867
  [hep-ph]} \BibitemShut {NoStop}%
\bibitem [{\citenamefont {Co}\ \emph {et~al.}(2018)\citenamefont {Co},
  \citenamefont {Hall},\ and\ \citenamefont {Harigaya}}]{Co:2017mop}%
  \BibitemOpen
  \bibfield  {author} {\bibinfo {author} {\bibfnamefont {R.~T.}\ \bibnamefont
  {Co}}, \bibinfo {author} {\bibfnamefont {L.~J.}\ \bibnamefont {Hall}}, \ and\
  \bibinfo {author} {\bibfnamefont {K.}~\bibnamefont {Harigaya}},\ }\href
  {\doibase 10.1103/PhysRevLett.120.211602} {\bibfield  {journal} {\bibinfo
  {journal} {Phys. Rev. Lett.}\ }\textbf {\bibinfo {volume} {120}},\ \bibinfo
  {pages} {211602} (\bibinfo {year} {2018})},\ \Eprint
  {http://arxiv.org/abs/1711.10486} {arXiv:1711.10486 [hep-ph]} \BibitemShut
  {NoStop}%
\bibitem [{\citenamefont {Blinov}\ \emph {et~al.}(2019)\citenamefont {Blinov},
  \citenamefont {Dolan}, \citenamefont {Draper},\ and\ \citenamefont
  {Kozaczuk}}]{Blinov:2019rhb}%
  \BibitemOpen
  \bibfield  {author} {\bibinfo {author} {\bibfnamefont {N.}~\bibnamefont
  {Blinov}}, \bibinfo {author} {\bibfnamefont {M.~J.}\ \bibnamefont {Dolan}},
  \bibinfo {author} {\bibfnamefont {P.}~\bibnamefont {Draper}}, \ and\ \bibinfo
  {author} {\bibfnamefont {J.}~\bibnamefont {Kozaczuk}},\ }\href {\doibase
  10.1103/PhysRevD.100.015049} {\bibfield  {journal} {\bibinfo  {journal}
  {Phys. Rev. D}\ }\textbf {\bibinfo {volume} {100}},\ \bibinfo {pages}
  {015049} (\bibinfo {year} {2019})},\ \Eprint
  {http://arxiv.org/abs/1905.06952} {arXiv:1905.06952 [hep-ph]} \BibitemShut
  {NoStop}%
\bibitem [{\citenamefont {Co}\ \emph {et~al.}(2020)\citenamefont {Co},
  \citenamefont {Hall},\ and\ \citenamefont {Harigaya}}]{Co:2019jts}%
  \BibitemOpen
  \bibfield  {author} {\bibinfo {author} {\bibfnamefont {R.~T.}\ \bibnamefont
  {Co}}, \bibinfo {author} {\bibfnamefont {L.~J.}\ \bibnamefont {Hall}}, \ and\
  \bibinfo {author} {\bibfnamefont {K.}~\bibnamefont {Harigaya}},\ }\href
  {\doibase 10.1103/PhysRevLett.124.251802} {\bibfield  {journal} {\bibinfo
  {journal} {Phys. Rev. Lett.}\ }\textbf {\bibinfo {volume} {124}},\ \bibinfo
  {pages} {251802} (\bibinfo {year} {2020})},\ \Eprint
  {http://arxiv.org/abs/1910.14152} {arXiv:1910.14152 [hep-ph]} \BibitemShut
  {NoStop}%
\bibitem [{\citenamefont {Co}\ \emph {et~al.}(2021)\citenamefont {Co},
  \citenamefont {Hall},\ and\ \citenamefont {Harigaya}}]{Co:2020xlh}%
  \BibitemOpen
  \bibfield  {author} {\bibinfo {author} {\bibfnamefont {R.~T.}\ \bibnamefont
  {Co}}, \bibinfo {author} {\bibfnamefont {L.~J.}\ \bibnamefont {Hall}}, \ and\
  \bibinfo {author} {\bibfnamefont {K.}~\bibnamefont {Harigaya}},\ }\href
  {\doibase 10.1007/JHEP01(2021)172} {\bibfield  {journal} {\bibinfo  {journal}
  {JHEP}\ }\textbf {\bibinfo {volume} {01}},\ \bibinfo {pages} {172} (\bibinfo
  {year} {2021})},\ \Eprint {http://arxiv.org/abs/2006.04809} {arXiv:2006.04809
  [hep-ph]} \BibitemShut {NoStop}%
\bibitem [{\citenamefont {Di~Luzio}\ \emph
  {et~al.}(2021{\natexlab{a}})\citenamefont {Di~Luzio}, \citenamefont {Gavela},
  \citenamefont {Quilez},\ and\ \citenamefont {Ringwald}}]{DiLuzio:2021gos}%
  \BibitemOpen
  \bibfield  {author} {\bibinfo {author} {\bibfnamefont {L.}~\bibnamefont
  {Di~Luzio}}, \bibinfo {author} {\bibfnamefont {B.}~\bibnamefont {Gavela}},
  \bibinfo {author} {\bibfnamefont {P.}~\bibnamefont {Quilez}}, \ and\ \bibinfo
  {author} {\bibfnamefont {A.}~\bibnamefont {Ringwald}},\ }\href {\doibase
  10.1088/1475-7516/2021/10/001} {\bibfield  {journal} {\bibinfo  {journal}
  {JCAP}\ }\textbf {\bibinfo {volume} {10}},\ \bibinfo {pages} {001} (\bibinfo
  {year} {2021}{\natexlab{a}})},\ \Eprint {http://arxiv.org/abs/2102.01082}
  {arXiv:2102.01082 [hep-ph]} \BibitemShut {NoStop}%
\bibitem [{\citenamefont {Takahashi}\ \emph {et~al.}(2018)\citenamefont
  {Takahashi}, \citenamefont {Yin},\ and\ \citenamefont
  {Guth}}]{Takahashi:2018tdu}%
  \BibitemOpen
  \bibfield  {author} {\bibinfo {author} {\bibfnamefont {F.}~\bibnamefont
  {Takahashi}}, \bibinfo {author} {\bibfnamefont {W.}~\bibnamefont {Yin}}, \
  and\ \bibinfo {author} {\bibfnamefont {A.~H.}\ \bibnamefont {Guth}},\ }\href
  {\doibase 10.1103/PhysRevD.98.015042} {\bibfield  {journal} {\bibinfo
  {journal} {Phys. Rev. D}\ }\textbf {\bibinfo {volume} {98}},\ \bibinfo
  {pages} {015042} (\bibinfo {year} {2018})},\ \Eprint
  {http://arxiv.org/abs/1805.08763} {arXiv:1805.08763 [hep-ph]} \BibitemShut
  {NoStop}%
\bibitem [{\citenamefont {Co}\ \emph {et~al.}(2019)\citenamefont {Co},
  \citenamefont {Gonzalez},\ and\ \citenamefont {Harigaya}}]{Co:2018mho}%
  \BibitemOpen
  \bibfield  {author} {\bibinfo {author} {\bibfnamefont {R.~T.}\ \bibnamefont
  {Co}}, \bibinfo {author} {\bibfnamefont {E.}~\bibnamefont {Gonzalez}}, \ and\
  \bibinfo {author} {\bibfnamefont {K.}~\bibnamefont {Harigaya}},\ }\href
  {\doibase 10.1007/JHEP05(2019)163} {\bibfield  {journal} {\bibinfo  {journal}
  {JHEP}\ }\textbf {\bibinfo {volume} {05}},\ \bibinfo {pages} {163} (\bibinfo
  {year} {2019})},\ \Eprint {http://arxiv.org/abs/1812.11192} {arXiv:1812.11192
  [hep-ph]} \BibitemShut {NoStop}%
\bibitem [{\citenamefont {Ho}\ \emph {et~al.}(2019)\citenamefont {Ho},
  \citenamefont {Takahashi},\ and\ \citenamefont {Yin}}]{Ho:2019ayl}%
  \BibitemOpen
  \bibfield  {author} {\bibinfo {author} {\bibfnamefont {S.-Y.}\ \bibnamefont
  {Ho}}, \bibinfo {author} {\bibfnamefont {F.}~\bibnamefont {Takahashi}}, \
  and\ \bibinfo {author} {\bibfnamefont {W.}~\bibnamefont {Yin}},\ }\href
  {\doibase 10.1007/JHEP04(2019)149} {\bibfield  {journal} {\bibinfo  {journal}
  {JHEP}\ }\textbf {\bibinfo {volume} {04}},\ \bibinfo {pages} {149} (\bibinfo
  {year} {2019})},\ \Eprint {http://arxiv.org/abs/1901.01240} {arXiv:1901.01240
  [hep-ph]} \BibitemShut {NoStop}%
\bibitem [{\citenamefont {Harigaya}\ and\ \citenamefont
  {Leedom}(2020)}]{Harigaya:2019qnl}%
  \BibitemOpen
  \bibfield  {author} {\bibinfo {author} {\bibfnamefont {K.}~\bibnamefont
  {Harigaya}}\ and\ \bibinfo {author} {\bibfnamefont {J.~M.}\ \bibnamefont
  {Leedom}},\ }\href {\doibase 10.1007/JHEP06(2020)034} {\bibfield  {journal}
  {\bibinfo  {journal} {JHEP}\ }\textbf {\bibinfo {volume} {06}},\ \bibinfo
  {pages} {034} (\bibinfo {year} {2020})},\ \Eprint
  {http://arxiv.org/abs/1910.04163} {arXiv:1910.04163 [hep-ph]} \BibitemShut
  {NoStop}%
\bibitem [{\citenamefont {Asadi}\ \emph {et~al.}(2022)\citenamefont {Asadi}
  \emph {et~al.}}]{Asadi:2022njl}%
  \BibitemOpen
  \bibfield  {author} {\bibinfo {author} {\bibfnamefont {P.}~\bibnamefont
  {Asadi}} \emph {et~al.},\ }\href@noop {} {\  (\bibinfo {year} {2022})},\
  \Eprint {http://arxiv.org/abs/2203.06680} {arXiv:2203.06680 [hep-ph]}
  \BibitemShut {NoStop}%
\bibitem [{\citenamefont {Kitano}\ \emph {et~al.}(2023)\citenamefont {Kitano},
  \citenamefont {Suzuki},\ and\ \citenamefont {Yin}}]{Kitano:2023mra}%
  \BibitemOpen
  \bibfield  {author} {\bibinfo {author} {\bibfnamefont {R.}~\bibnamefont
  {Kitano}}, \bibinfo {author} {\bibfnamefont {M.}~\bibnamefont {Suzuki}}, \
  and\ \bibinfo {author} {\bibfnamefont {W.}~\bibnamefont {Yin}},\ }\href
  {\doibase 10.1007/JHEP11(2023)103} {\bibfield  {journal} {\bibinfo  {journal}
  {JHEP}\ }\textbf {\bibinfo {volume} {11}},\ \bibinfo {pages} {103} (\bibinfo
  {year} {2023})},\ \Eprint {http://arxiv.org/abs/2307.15059} {arXiv:2307.15059
  [hep-ph]} \BibitemShut {NoStop}%
\bibitem [{\citenamefont {Graham}\ \emph {et~al.}(2016)\citenamefont {Graham},
  \citenamefont {Mardon},\ and\ \citenamefont {Rajendran}}]{Graham:2015rva}%
  \BibitemOpen
  \bibfield  {author} {\bibinfo {author} {\bibfnamefont {P.~W.}\ \bibnamefont
  {Graham}}, \bibinfo {author} {\bibfnamefont {J.}~\bibnamefont {Mardon}}, \
  and\ \bibinfo {author} {\bibfnamefont {S.}~\bibnamefont {Rajendran}},\ }\href
  {\doibase 10.1103/PhysRevD.93.103520} {\bibfield  {journal} {\bibinfo
  {journal} {Phys. Rev. D}\ }\textbf {\bibinfo {volume} {93}},\ \bibinfo
  {pages} {103520} (\bibinfo {year} {2016})},\ \Eprint
  {http://arxiv.org/abs/1504.02102} {arXiv:1504.02102 [hep-ph]} \BibitemShut
  {NoStop}%
\bibitem [{\citenamefont {Redi}\ and\ \citenamefont
  {Tesi}(2023)}]{Redi:2022llj}%
  \BibitemOpen
  \bibfield  {author} {\bibinfo {author} {\bibfnamefont {M.}~\bibnamefont
  {Redi}}\ and\ \bibinfo {author} {\bibfnamefont {A.}~\bibnamefont {Tesi}},\
  }\href {\doibase 10.1103/PhysRevD.107.095032} {\bibfield  {journal} {\bibinfo
   {journal} {Phys. Rev. D}\ }\textbf {\bibinfo {volume} {107}},\ \bibinfo
  {pages} {095032} (\bibinfo {year} {2023})},\ \Eprint
  {http://arxiv.org/abs/2211.06421} {arXiv:2211.06421 [hep-ph]} \BibitemShut
  {NoStop}%
\bibitem [{\citenamefont {Nakai}\ \emph {et~al.}(2020)\citenamefont {Nakai},
  \citenamefont {Namba},\ and\ \citenamefont {Wang}}]{Nakai:2020cfw}%
  \BibitemOpen
  \bibfield  {author} {\bibinfo {author} {\bibfnamefont {Y.}~\bibnamefont
  {Nakai}}, \bibinfo {author} {\bibfnamefont {R.}~\bibnamefont {Namba}}, \ and\
  \bibinfo {author} {\bibfnamefont {Z.}~\bibnamefont {Wang}},\ }\href {\doibase
  10.1007/JHEP12(2020)170} {\bibfield  {journal} {\bibinfo  {journal} {JHEP}\
  }\textbf {\bibinfo {volume} {12}},\ \bibinfo {pages} {170} (\bibinfo {year}
  {2020})},\ \Eprint {http://arxiv.org/abs/2004.10743} {arXiv:2004.10743
  [hep-ph]} \BibitemShut {NoStop}%
\bibitem [{\citenamefont {Ellis}\ \emph {et~al.}(1984)\citenamefont {Ellis},
  \citenamefont {Lahanas}, \citenamefont {Nanopoulos},\ and\ \citenamefont
  {Tamvakis}}]{Ellis:1983sf}%
  \BibitemOpen
  \bibfield  {author} {\bibinfo {author} {\bibfnamefont {J.~R.}\ \bibnamefont
  {Ellis}}, \bibinfo {author} {\bibfnamefont {A.~B.}\ \bibnamefont {Lahanas}},
  \bibinfo {author} {\bibfnamefont {D.~V.}\ \bibnamefont {Nanopoulos}}, \ and\
  \bibinfo {author} {\bibfnamefont {K.}~\bibnamefont {Tamvakis}},\ }\href
  {\doibase 10.1016/0370-2693(84)91378-9} {\bibfield  {journal} {\bibinfo
  {journal} {Phys. Lett. B}\ }\textbf {\bibinfo {volume} {134}},\ \bibinfo
  {pages} {429} (\bibinfo {year} {1984})}\BibitemShut {NoStop}%
\bibitem [{\citenamefont {Ellis}\ \emph {et~al.}(2013)\citenamefont {Ellis},
  \citenamefont {Nanopoulos},\ and\ \citenamefont {Olive}}]{Ellis:2013xoa}%
  \BibitemOpen
  \bibfield  {author} {\bibinfo {author} {\bibfnamefont {J.}~\bibnamefont
  {Ellis}}, \bibinfo {author} {\bibfnamefont {D.~V.}\ \bibnamefont
  {Nanopoulos}}, \ and\ \bibinfo {author} {\bibfnamefont {K.~A.}\ \bibnamefont
  {Olive}},\ }\href {\doibase 10.1103/PhysRevLett.111.111301} {\bibfield
  {journal} {\bibinfo  {journal} {Phys. Rev. Lett.}\ }\textbf {\bibinfo
  {volume} {111}},\ \bibinfo {pages} {111301} (\bibinfo {year} {2013})},\
  \bibinfo {note} {[Erratum: Phys.Rev.Lett. 111, 129902 (2013)]},\ \Eprint
  {http://arxiv.org/abs/1305.1247} {arXiv:1305.1247 [hep-th]} \BibitemShut
  {NoStop}%
\bibitem [{\citenamefont {Kobayashi}\ \emph {et~al.}(2016)\citenamefont
  {Kobayashi}, \citenamefont {Nitta},\ and\ \citenamefont
  {Urakawa}}]{Kobayashi:2016mzg}%
  \BibitemOpen
  \bibfield  {author} {\bibinfo {author} {\bibfnamefont {T.}~\bibnamefont
  {Kobayashi}}, \bibinfo {author} {\bibfnamefont {D.}~\bibnamefont {Nitta}}, \
  and\ \bibinfo {author} {\bibfnamefont {Y.}~\bibnamefont {Urakawa}},\ }\href
  {\doibase 10.1088/1475-7516/2016/08/014} {\bibfield  {journal} {\bibinfo
  {journal} {JCAP}\ }\textbf {\bibinfo {volume} {08}},\ \bibinfo {pages} {014}
  (\bibinfo {year} {2016})},\ \Eprint {http://arxiv.org/abs/1604.02995}
  {arXiv:1604.02995 [hep-th]} \BibitemShut {NoStop}%
\bibitem [{\citenamefont {Abe}\ \emph {et~al.}(2023)\citenamefont {Abe},
  \citenamefont {Higaki}, \citenamefont {Kaneko}, \citenamefont {Kobayashi},\
  and\ \citenamefont {Otsuka}}]{Abe:2023ylh}%
  \BibitemOpen
  \bibfield  {author} {\bibinfo {author} {\bibfnamefont {Y.}~\bibnamefont
  {Abe}}, \bibinfo {author} {\bibfnamefont {T.}~\bibnamefont {Higaki}},
  \bibinfo {author} {\bibfnamefont {F.}~\bibnamefont {Kaneko}}, \bibinfo
  {author} {\bibfnamefont {T.}~\bibnamefont {Kobayashi}}, \ and\ \bibinfo
  {author} {\bibfnamefont {H.}~\bibnamefont {Otsuka}},\ }\href {\doibase
  10.1007/JHEP06(2023)187} {\bibfield  {journal} {\bibinfo  {journal} {JHEP}\
  }\textbf {\bibinfo {volume} {06}},\ \bibinfo {pages} {187} (\bibinfo {year}
  {2023})},\ \Eprint {http://arxiv.org/abs/2303.02947} {arXiv:2303.02947
  [hep-ph]} \BibitemShut {NoStop}%
\bibitem [{\citenamefont {Gibbons}\ and\ \citenamefont
  {Hawking}(1977)}]{Gibbons:1977mu}%
  \BibitemOpen
  \bibfield  {author} {\bibinfo {author} {\bibfnamefont {G.~W.}\ \bibnamefont
  {Gibbons}}\ and\ \bibinfo {author} {\bibfnamefont {S.~W.}\ \bibnamefont
  {Hawking}},\ }\href {\doibase 10.1103/PhysRevD.15.2738} {\bibfield  {journal}
  {\bibinfo  {journal} {Phys. Rev. D}\ }\textbf {\bibinfo {volume} {15}},\
  \bibinfo {pages} {2738} (\bibinfo {year} {1977})}\BibitemShut {NoStop}%
\bibitem [{\citenamefont {Aghanim}\ \emph {et~al.}(2020)\citenamefont {Aghanim}
  \emph {et~al.}}]{Planck:2018vyg}%
  \BibitemOpen
  \bibfield  {author} {\bibinfo {author} {\bibfnamefont {N.}~\bibnamefont
  {Aghanim}} \emph {et~al.} (\bibinfo {collaboration} {Planck}),\ }\href
  {\doibase 10.1051/0004-6361/201833910} {\bibfield  {journal} {\bibinfo
  {journal} {Astron. Astrophys.}\ }\textbf {\bibinfo {volume} {641}},\ \bibinfo
  {pages} {A6} (\bibinfo {year} {2020})},\ \bibinfo {note} {[Erratum:
  Astron.Astrophys. 652, C4 (2021)]},\ \Eprint
  {http://arxiv.org/abs/1807.06209} {arXiv:1807.06209 [astro-ph.CO]}
  \BibitemShut {NoStop}%
\bibitem [{\citenamefont {Bartolo}\ \emph {et~al.}(2013)\citenamefont
  {Bartolo}, \citenamefont {Matarrese}, \citenamefont {Peloso},\ and\
  \citenamefont {Ricciardone}}]{Bartolo:2012sd}%
  \BibitemOpen
  \bibfield  {author} {\bibinfo {author} {\bibfnamefont {N.}~\bibnamefont
  {Bartolo}}, \bibinfo {author} {\bibfnamefont {S.}~\bibnamefont {Matarrese}},
  \bibinfo {author} {\bibfnamefont {M.}~\bibnamefont {Peloso}}, \ and\ \bibinfo
  {author} {\bibfnamefont {A.}~\bibnamefont {Ricciardone}},\ }\href {\doibase
  10.1103/PhysRevD.87.023504} {\bibfield  {journal} {\bibinfo  {journal} {Phys.
  Rev. D}\ }\textbf {\bibinfo {volume} {87}},\ \bibinfo {pages} {023504}
  (\bibinfo {year} {2013})},\ \Eprint {http://arxiv.org/abs/1210.3257}
  {arXiv:1210.3257 [astro-ph.CO]} \BibitemShut {NoStop}%
\bibitem [{\citenamefont {Ade}\ \emph {et~al.}(2021)\citenamefont {Ade} \emph
  {et~al.}}]{BICEP:2021xfz}%
  \BibitemOpen
  \bibfield  {author} {\bibinfo {author} {\bibfnamefont {P.~A.~R.}\
  \bibnamefont {Ade}} \emph {et~al.} (\bibinfo {collaboration} {BICEP, Keck}),\
  }\href {\doibase 10.1103/PhysRevLett.127.151301} {\bibfield  {journal}
  {\bibinfo  {journal} {Phys. Rev. Lett.}\ }\textbf {\bibinfo {volume} {127}},\
  \bibinfo {pages} {151301} (\bibinfo {year} {2021})},\ \Eprint
  {http://arxiv.org/abs/2110.00483} {arXiv:2110.00483 [astro-ph.CO]}
  \BibitemShut {NoStop}%
\bibitem [{\citenamefont {Feix}\ \emph {et~al.}(2019)\citenamefont {Feix},
  \citenamefont {Frank}, \citenamefont {Pargner}, \citenamefont {Reischke},
  \citenamefont {Sch\"afer},\ and\ \citenamefont {Schwetz}}]{Feix:2019lpo}%
  \BibitemOpen
  \bibfield  {author} {\bibinfo {author} {\bibfnamefont {M.}~\bibnamefont
  {Feix}}, \bibinfo {author} {\bibfnamefont {J.}~\bibnamefont {Frank}},
  \bibinfo {author} {\bibfnamefont {A.}~\bibnamefont {Pargner}}, \bibinfo
  {author} {\bibfnamefont {R.}~\bibnamefont {Reischke}}, \bibinfo {author}
  {\bibfnamefont {B.~M.}\ \bibnamefont {Sch\"afer}}, \ and\ \bibinfo {author}
  {\bibfnamefont {T.}~\bibnamefont {Schwetz}},\ }\href {\doibase
  10.1088/1475-7516/2019/05/021} {\bibfield  {journal} {\bibinfo  {journal}
  {JCAP}\ }\textbf {\bibinfo {volume} {05}},\ \bibinfo {pages} {021} (\bibinfo
  {year} {2019})},\ \Eprint {http://arxiv.org/abs/1903.06194} {arXiv:1903.06194
  [astro-ph.CO]} \BibitemShut {NoStop}%
\bibitem [{\citenamefont {Andr\'e}\ \emph {et~al.}(2014)\citenamefont {Andr\'e}
  \emph {et~al.}}]{PRISM:2013fvg}%
  \BibitemOpen
  \bibfield  {author} {\bibinfo {author} {\bibfnamefont {P.}~\bibnamefont
  {Andr\'e}} \emph {et~al.} (\bibinfo {collaboration} {PRISM}),\ }\href
  {\doibase 10.1088/1475-7516/2014/02/006} {\bibfield  {journal} {\bibinfo
  {journal} {JCAP}\ }\textbf {\bibinfo {volume} {02}},\ \bibinfo {pages} {006}
  (\bibinfo {year} {2014})},\ \Eprint {http://arxiv.org/abs/1310.1554}
  {arXiv:1310.1554 [astro-ph.CO]} \BibitemShut {NoStop}%
\bibitem [{\citenamefont {Abazajian}\ \emph {et~al.}(2016)\citenamefont
  {Abazajian} \emph {et~al.}}]{CMB-S4:2016ple}%
  \BibitemOpen
  \bibfield  {author} {\bibinfo {author} {\bibfnamefont {K.~N.}\ \bibnamefont
  {Abazajian}} \emph {et~al.} (\bibinfo {collaboration} {CMB-S4}),\ }\href@noop
  {} {\  (\bibinfo {year} {2016})},\ \Eprint {http://arxiv.org/abs/1610.02743}
  {arXiv:1610.02743 [astro-ph.CO]} \BibitemShut {NoStop}%
\bibitem [{\citenamefont {Pritchard}\ \emph {et~al.}(2015)\citenamefont
  {Pritchard} \emph {et~al.}}]{Cosmology-SWG:2015tjb}%
  \BibitemOpen
  \bibfield  {author} {\bibinfo {author} {\bibfnamefont {J.}~\bibnamefont
  {Pritchard}} \emph {et~al.} (\bibinfo {collaboration} {Cosmology-SWG,
  EoR/CD-SWG}),\ }\href {\doibase 10.22323/1.215.0012} {\bibfield  {journal}
  {\bibinfo  {journal} {PoS}\ }\textbf {\bibinfo {volume} {AASKA14}},\ \bibinfo
  {pages} {012} (\bibinfo {year} {2015})},\ \Eprint
  {http://arxiv.org/abs/1501.04291} {arXiv:1501.04291 [astro-ph.CO]}
  \BibitemShut {NoStop}%
\bibitem [{\citenamefont {Weltman}\ \emph {et~al.}(2020)\citenamefont {Weltman}
  \emph {et~al.}}]{Weltman:2018zrl}%
  \BibitemOpen
  \bibfield  {author} {\bibinfo {author} {\bibfnamefont {A.}~\bibnamefont
  {Weltman}} \emph {et~al.},\ }\href {\doibase 10.1017/pasa.2019.42} {\bibfield
   {journal} {\bibinfo  {journal} {Publ. Astron. Soc. Austral.}\ }\textbf
  {\bibinfo {volume} {37}},\ \bibinfo {pages} {e002} (\bibinfo {year}
  {2020})},\ \Eprint {http://arxiv.org/abs/1810.02680} {arXiv:1810.02680
  [astro-ph.CO]} \BibitemShut {NoStop}%
\bibitem [{\citenamefont {Reyn\'es}\ \emph {et~al.}(2021)\citenamefont
  {Reyn\'es}, \citenamefont {Matthews}, \citenamefont {Reynolds}, \citenamefont
  {Russell}, \citenamefont {Smith},\ and\ \citenamefont
  {Marsh}}]{Reynes:2021bpe}%
  \BibitemOpen
  \bibfield  {author} {\bibinfo {author} {\bibfnamefont {J.~S.}\ \bibnamefont
  {Reyn\'es}}, \bibinfo {author} {\bibfnamefont {J.~H.}\ \bibnamefont
  {Matthews}}, \bibinfo {author} {\bibfnamefont {C.~S.}\ \bibnamefont
  {Reynolds}}, \bibinfo {author} {\bibfnamefont {H.~R.}\ \bibnamefont
  {Russell}}, \bibinfo {author} {\bibfnamefont {R.~N.}\ \bibnamefont {Smith}},
  \ and\ \bibinfo {author} {\bibfnamefont {M.~C.~D.}\ \bibnamefont {Marsh}},\
  }\href {\doibase 10.1093/mnras/stab3464} {\bibfield  {journal} {\bibinfo
  {journal} {Mon. Not. Roy. Astron. Soc.}\ }\textbf {\bibinfo {volume} {510}},\
  \bibinfo {pages} {1264} (\bibinfo {year} {2021})},\ \Eprint
  {http://arxiv.org/abs/2109.03261} {arXiv:2109.03261 [astro-ph.HE]}
  \BibitemShut {NoStop}%
\bibitem [{\citenamefont {Reynolds}\ \emph {et~al.}(2020)\citenamefont
  {Reynolds}, \citenamefont {Marsh}, \citenamefont {Russell}, \citenamefont
  {Fabian}, \citenamefont {Smith}, \citenamefont {Tombesi},\ and\ \citenamefont
  {Veilleux}}]{Reynolds:2019uqt}%
  \BibitemOpen
  \bibfield  {author} {\bibinfo {author} {\bibfnamefont {C.~S.}\ \bibnamefont
  {Reynolds}}, \bibinfo {author} {\bibfnamefont {M.~C.~D.}\ \bibnamefont
  {Marsh}}, \bibinfo {author} {\bibfnamefont {H.~R.}\ \bibnamefont {Russell}},
  \bibinfo {author} {\bibfnamefont {A.~C.}\ \bibnamefont {Fabian}}, \bibinfo
  {author} {\bibfnamefont {R.}~\bibnamefont {Smith}}, \bibinfo {author}
  {\bibfnamefont {F.}~\bibnamefont {Tombesi}}, \ and\ \bibinfo {author}
  {\bibfnamefont {S.}~\bibnamefont {Veilleux}},\ }\href {\doibase
  10.3847/1538-4357/ab6a0c} {\bibfield  {journal} {\bibinfo  {journal}
  {Astrophys. J.}\ }\textbf {\bibinfo {volume} {890}},\ \bibinfo {pages} {59}
  (\bibinfo {year} {2020})},\ \Eprint {http://arxiv.org/abs/1907.05475}
  {arXiv:1907.05475 [hep-ph]} \BibitemShut {NoStop}%
\bibitem [{\citenamefont {Marsh}\ \emph {et~al.}(2017)\citenamefont {Marsh},
  \citenamefont {Russell}, \citenamefont {Fabian}, \citenamefont {McNamara},
  \citenamefont {Nulsen},\ and\ \citenamefont {Reynolds}}]{Marsh:2017yvc}%
  \BibitemOpen
  \bibfield  {author} {\bibinfo {author} {\bibfnamefont {M.~C.~D.}\
  \bibnamefont {Marsh}}, \bibinfo {author} {\bibfnamefont {H.~R.}\ \bibnamefont
  {Russell}}, \bibinfo {author} {\bibfnamefont {A.~C.}\ \bibnamefont {Fabian}},
  \bibinfo {author} {\bibfnamefont {B.~P.}\ \bibnamefont {McNamara}}, \bibinfo
  {author} {\bibfnamefont {P.}~\bibnamefont {Nulsen}}, \ and\ \bibinfo {author}
  {\bibfnamefont {C.~S.}\ \bibnamefont {Reynolds}},\ }\href {\doibase
  10.1088/1475-7516/2017/12/036} {\bibfield  {journal} {\bibinfo  {journal}
  {JCAP}\ }\textbf {\bibinfo {volume} {12}},\ \bibinfo {pages} {036} (\bibinfo
  {year} {2017})},\ \Eprint {http://arxiv.org/abs/1703.07354} {arXiv:1703.07354
  [hep-ph]} \BibitemShut {NoStop}%
\bibitem [{\citenamefont {Dessert}\ \emph {et~al.}(2020)\citenamefont
  {Dessert}, \citenamefont {Foster},\ and\ \citenamefont
  {Safdi}}]{Dessert:2020lil}%
  \BibitemOpen
  \bibfield  {author} {\bibinfo {author} {\bibfnamefont {C.}~\bibnamefont
  {Dessert}}, \bibinfo {author} {\bibfnamefont {J.~W.}\ \bibnamefont {Foster}},
  \ and\ \bibinfo {author} {\bibfnamefont {B.~R.}\ \bibnamefont {Safdi}},\
  }\href {\doibase 10.1103/PhysRevLett.125.261102} {\bibfield  {journal}
  {\bibinfo  {journal} {Phys. Rev. Lett.}\ }\textbf {\bibinfo {volume} {125}},\
  \bibinfo {pages} {261102} (\bibinfo {year} {2020})},\ \Eprint
  {http://arxiv.org/abs/2008.03305} {arXiv:2008.03305 [hep-ph]} \BibitemShut
  {NoStop}%
\bibitem [{\citenamefont {Hoof}\ and\ \citenamefont
  {Schulz}(2023)}]{Hoof:2022xbe}%
  \BibitemOpen
  \bibfield  {author} {\bibinfo {author} {\bibfnamefont {S.}~\bibnamefont
  {Hoof}}\ and\ \bibinfo {author} {\bibfnamefont {L.}~\bibnamefont {Schulz}},\
  }\href {\doibase 10.1088/1475-7516/2023/03/054} {\bibfield  {journal}
  {\bibinfo  {journal} {JCAP}\ }\textbf {\bibinfo {volume} {03}},\ \bibinfo
  {pages} {054} (\bibinfo {year} {2023})},\ \Eprint
  {http://arxiv.org/abs/2212.09764} {arXiv:2212.09764 [hep-ph]} \BibitemShut
  {NoStop}%
\bibitem [{\citenamefont {Ajello}\ \emph {et~al.}(2016)\citenamefont {Ajello}
  \emph {et~al.}}]{Fermi-LAT:2016nkz}%
  \BibitemOpen
  \bibfield  {author} {\bibinfo {author} {\bibfnamefont {M.}~\bibnamefont
  {Ajello}} \emph {et~al.} (\bibinfo {collaboration} {Fermi-LAT}),\ }\href
  {\doibase 10.1103/PhysRevLett.116.161101} {\bibfield  {journal} {\bibinfo
  {journal} {Phys. Rev. Lett.}\ }\textbf {\bibinfo {volume} {116}},\ \bibinfo
  {pages} {161101} (\bibinfo {year} {2016})},\ \Eprint
  {http://arxiv.org/abs/1603.06978} {arXiv:1603.06978 [astro-ph.HE]}
  \BibitemShut {NoStop}%
\bibitem [{\citenamefont {Dessert}\ \emph {et~al.}(2022)\citenamefont
  {Dessert}, \citenamefont {Dunsky},\ and\ \citenamefont
  {Safdi}}]{Dessert:2022yqq}%
  \BibitemOpen
  \bibfield  {author} {\bibinfo {author} {\bibfnamefont {C.}~\bibnamefont
  {Dessert}}, \bibinfo {author} {\bibfnamefont {D.}~\bibnamefont {Dunsky}}, \
  and\ \bibinfo {author} {\bibfnamefont {B.~R.}\ \bibnamefont {Safdi}},\ }\href
  {\doibase 10.1103/PhysRevD.105.103034} {\bibfield  {journal} {\bibinfo
  {journal} {Phys. Rev. D}\ }\textbf {\bibinfo {volume} {105}},\ \bibinfo
  {pages} {103034} (\bibinfo {year} {2022})},\ \Eprint
  {http://arxiv.org/abs/2203.04319} {arXiv:2203.04319 [hep-ph]} \BibitemShut
  {NoStop}%
\bibitem [{\citenamefont {Noordhuis}\ \emph {et~al.}(2023)\citenamefont
  {Noordhuis}, \citenamefont {Prabhu}, \citenamefont {Witte}, \citenamefont
  {Chen}, \citenamefont {Cruz},\ and\ \citenamefont
  {Weniger}}]{Noordhuis:2022ljw}%
  \BibitemOpen
  \bibfield  {author} {\bibinfo {author} {\bibfnamefont {D.}~\bibnamefont
  {Noordhuis}}, \bibinfo {author} {\bibfnamefont {A.}~\bibnamefont {Prabhu}},
  \bibinfo {author} {\bibfnamefont {S.~J.}\ \bibnamefont {Witte}}, \bibinfo
  {author} {\bibfnamefont {A.~Y.}\ \bibnamefont {Chen}}, \bibinfo {author}
  {\bibfnamefont {F.}~\bibnamefont {Cruz}}, \ and\ \bibinfo {author}
  {\bibfnamefont {C.}~\bibnamefont {Weniger}},\ }\href {\doibase
  10.1103/PhysRevLett.131.111004} {\bibfield  {journal} {\bibinfo  {journal}
  {Phys. Rev. Lett.}\ }\textbf {\bibinfo {volume} {131}},\ \bibinfo {pages}
  {111004} (\bibinfo {year} {2023})},\ \Eprint
  {http://arxiv.org/abs/2209.09917} {arXiv:2209.09917 [hep-ph]} \BibitemShut
  {NoStop}%
\bibitem [{\citenamefont {Ayala}\ \emph {et~al.}(2014)\citenamefont {Ayala},
  \citenamefont {Dom\'\i{}nguez}, \citenamefont {Giannotti}, \citenamefont
  {Mirizzi},\ and\ \citenamefont {Straniero}}]{Ayala:2014pea}%
  \BibitemOpen
  \bibfield  {author} {\bibinfo {author} {\bibfnamefont {A.}~\bibnamefont
  {Ayala}}, \bibinfo {author} {\bibfnamefont {I.}~\bibnamefont
  {Dom\'\i{}nguez}}, \bibinfo {author} {\bibfnamefont {M.}~\bibnamefont
  {Giannotti}}, \bibinfo {author} {\bibfnamefont {A.}~\bibnamefont {Mirizzi}},
  \ and\ \bibinfo {author} {\bibfnamefont {O.}~\bibnamefont {Straniero}},\
  }\href {\doibase 10.1103/PhysRevLett.113.191302} {\bibfield  {journal}
  {\bibinfo  {journal} {Phys. Rev. Lett.}\ }\textbf {\bibinfo {volume} {113}},\
  \bibinfo {pages} {191302} (\bibinfo {year} {2014})},\ \Eprint
  {http://arxiv.org/abs/1406.6053} {arXiv:1406.6053 [astro-ph.SR]} \BibitemShut
  {NoStop}%
\bibitem [{\citenamefont {Balkin}\ \emph {et~al.}(2020)\citenamefont {Balkin},
  \citenamefont {Serra}, \citenamefont {Springmann},\ and\ \citenamefont
  {Weiler}}]{Balkin:2020dsr}%
  \BibitemOpen
  \bibfield  {author} {\bibinfo {author} {\bibfnamefont {R.}~\bibnamefont
  {Balkin}}, \bibinfo {author} {\bibfnamefont {J.}~\bibnamefont {Serra}},
  \bibinfo {author} {\bibfnamefont {K.}~\bibnamefont {Springmann}}, \ and\
  \bibinfo {author} {\bibfnamefont {A.}~\bibnamefont {Weiler}},\ }\href
  {\doibase 10.1007/JHEP07(2020)221} {\bibfield  {journal} {\bibinfo  {journal}
  {JHEP}\ }\textbf {\bibinfo {volume} {07}},\ \bibinfo {pages} {221} (\bibinfo
  {year} {2020})},\ \Eprint {http://arxiv.org/abs/2003.04903} {arXiv:2003.04903
  [hep-ph]} \BibitemShut {NoStop}%
\bibitem [{\citenamefont {Asztalos}\ \emph {et~al.}(2010)\citenamefont
  {Asztalos} \emph {et~al.}}]{ADMX:2009iij}%
  \BibitemOpen
  \bibfield  {author} {\bibinfo {author} {\bibfnamefont {S.~J.}\ \bibnamefont
  {Asztalos}} \emph {et~al.} (\bibinfo {collaboration} {ADMX}),\ }\href
  {\doibase 10.1103/PhysRevLett.104.041301} {\bibfield  {journal} {\bibinfo
  {journal} {Phys. Rev. Lett.}\ }\textbf {\bibinfo {volume} {104}},\ \bibinfo
  {pages} {041301} (\bibinfo {year} {2010})},\ \Eprint
  {http://arxiv.org/abs/0910.5914} {arXiv:0910.5914 [astro-ph.CO]} \BibitemShut
  {NoStop}%
\bibitem [{\citenamefont {Du}\ \emph {et~al.}(2018)\citenamefont {Du} \emph
  {et~al.}}]{ADMX:2018gho}%
  \BibitemOpen
  \bibfield  {author} {\bibinfo {author} {\bibfnamefont {N.}~\bibnamefont {Du}}
  \emph {et~al.} (\bibinfo {collaboration} {ADMX}),\ }\href {\doibase
  10.1103/PhysRevLett.120.151301} {\bibfield  {journal} {\bibinfo  {journal}
  {Phys. Rev. Lett.}\ }\textbf {\bibinfo {volume} {120}},\ \bibinfo {pages}
  {151301} (\bibinfo {year} {2018})},\ \Eprint
  {http://arxiv.org/abs/1804.05750} {arXiv:1804.05750 [hep-ex]} \BibitemShut
  {NoStop}%
\bibitem [{\citenamefont {Braine}\ \emph {et~al.}(2020)\citenamefont {Braine}
  \emph {et~al.}}]{ADMX:2019uok}%
  \BibitemOpen
  \bibfield  {author} {\bibinfo {author} {\bibfnamefont {T.}~\bibnamefont
  {Braine}} \emph {et~al.} (\bibinfo {collaboration} {ADMX}),\ }\href {\doibase
  10.1103/PhysRevLett.124.101303} {\bibfield  {journal} {\bibinfo  {journal}
  {Phys. Rev. Lett.}\ }\textbf {\bibinfo {volume} {124}},\ \bibinfo {pages}
  {101303} (\bibinfo {year} {2020})},\ \Eprint
  {http://arxiv.org/abs/1910.08638} {arXiv:1910.08638 [hep-ex]} \BibitemShut
  {NoStop}%
\bibitem [{\citenamefont {Bartram}\ \emph {et~al.}(2021)\citenamefont {Bartram}
  \emph {et~al.}}]{ADMX:2021nhd}%
  \BibitemOpen
  \bibfield  {author} {\bibinfo {author} {\bibfnamefont {C.}~\bibnamefont
  {Bartram}} \emph {et~al.} (\bibinfo {collaboration} {ADMX}),\ }\href
  {\doibase 10.1103/PhysRevLett.127.261803} {\bibfield  {journal} {\bibinfo
  {journal} {Phys. Rev. Lett.}\ }\textbf {\bibinfo {volume} {127}},\ \bibinfo
  {pages} {261803} (\bibinfo {year} {2021})},\ \Eprint
  {http://arxiv.org/abs/2110.06096} {arXiv:2110.06096 [hep-ex]} \BibitemShut
  {NoStop}%
\bibitem [{\citenamefont {De~Panfilis}\ \emph {et~al.}(1987)\citenamefont
  {De~Panfilis}, \citenamefont {Melissinos}, \citenamefont {Moskowitz},
  \citenamefont {Rogers}, \citenamefont {Semertzidis}, \citenamefont {Wuensch},
  \citenamefont {Halama}, \citenamefont {Prodell}, \citenamefont {Fowler},\
  and\ \citenamefont {Nezrick}}]{DePanfilis:1987dk}%
  \BibitemOpen
  \bibfield  {author} {\bibinfo {author} {\bibfnamefont {S.}~\bibnamefont
  {De~Panfilis}}, \bibinfo {author} {\bibfnamefont {A.~C.}\ \bibnamefont
  {Melissinos}}, \bibinfo {author} {\bibfnamefont {B.~E.}\ \bibnamefont
  {Moskowitz}}, \bibinfo {author} {\bibfnamefont {J.~T.}\ \bibnamefont
  {Rogers}}, \bibinfo {author} {\bibfnamefont {Y.~K.}\ \bibnamefont
  {Semertzidis}}, \bibinfo {author} {\bibfnamefont {W.}~\bibnamefont
  {Wuensch}}, \bibinfo {author} {\bibfnamefont {H.~J.}\ \bibnamefont {Halama}},
  \bibinfo {author} {\bibfnamefont {A.~G.}\ \bibnamefont {Prodell}}, \bibinfo
  {author} {\bibfnamefont {W.~B.}\ \bibnamefont {Fowler}}, \ and\ \bibinfo
  {author} {\bibfnamefont {F.~A.}\ \bibnamefont {Nezrick}},\ }\href {\doibase
  10.1103/PhysRevLett.59.839} {\bibfield  {journal} {\bibinfo  {journal} {Phys.
  Rev. Lett.}\ }\textbf {\bibinfo {volume} {59}},\ \bibinfo {pages} {839}
  (\bibinfo {year} {1987})}\BibitemShut {NoStop}%
\bibitem [{\citenamefont {Hagmann}\ \emph {et~al.}(1990)\citenamefont
  {Hagmann}, \citenamefont {Sikivie}, \citenamefont {Sullivan},\ and\
  \citenamefont {Tanner}}]{Hagmann:1990tj}%
  \BibitemOpen
  \bibfield  {author} {\bibinfo {author} {\bibfnamefont {C.}~\bibnamefont
  {Hagmann}}, \bibinfo {author} {\bibfnamefont {P.}~\bibnamefont {Sikivie}},
  \bibinfo {author} {\bibfnamefont {N.~S.}\ \bibnamefont {Sullivan}}, \ and\
  \bibinfo {author} {\bibfnamefont {D.~B.}\ \bibnamefont {Tanner}},\ }\href
  {\doibase 10.1103/PhysRevD.42.1297} {\bibfield  {journal} {\bibinfo
  {journal} {Phys. Rev. D}\ }\textbf {\bibinfo {volume} {42}},\ \bibinfo
  {pages} {1297} (\bibinfo {year} {1990})}\BibitemShut {NoStop}%
\bibitem [{\citenamefont {Zhong}\ \emph {et~al.}(2018)\citenamefont {Zhong}
  \emph {et~al.}}]{HAYSTAC:2018rwy}%
  \BibitemOpen
  \bibfield  {author} {\bibinfo {author} {\bibfnamefont {L.}~\bibnamefont
  {Zhong}} \emph {et~al.} (\bibinfo {collaboration} {HAYSTAC}),\ }\href
  {\doibase 10.1103/PhysRevD.97.092001} {\bibfield  {journal} {\bibinfo
  {journal} {Phys. Rev. D}\ }\textbf {\bibinfo {volume} {97}},\ \bibinfo
  {pages} {092001} (\bibinfo {year} {2018})},\ \Eprint
  {http://arxiv.org/abs/1803.03690} {arXiv:1803.03690 [hep-ex]} \BibitemShut
  {NoStop}%
\bibitem [{\citenamefont {Backes}\ \emph {et~al.}(2021)\citenamefont {Backes}
  \emph {et~al.}}]{HAYSTAC:2020kwv}%
  \BibitemOpen
  \bibfield  {author} {\bibinfo {author} {\bibfnamefont {K.~M.}\ \bibnamefont
  {Backes}} \emph {et~al.} (\bibinfo {collaboration} {HAYSTAC}),\ }\href
  {\doibase 10.1038/s41586-021-03226-7} {\bibfield  {journal} {\bibinfo
  {journal} {Nature}\ }\textbf {\bibinfo {volume} {590}},\ \bibinfo {pages}
  {238} (\bibinfo {year} {2021})},\ \Eprint {http://arxiv.org/abs/2008.01853}
  {arXiv:2008.01853 [quant-ph]} \BibitemShut {NoStop}%
\bibitem [{\citenamefont {Jewell}\ \emph {et~al.}(2023)\citenamefont {Jewell}
  \emph {et~al.}}]{HAYSTAC:2023cam}%
  \BibitemOpen
  \bibfield  {author} {\bibinfo {author} {\bibfnamefont {M.~J.}\ \bibnamefont
  {Jewell}} \emph {et~al.} (\bibinfo {collaboration} {HAYSTAC}),\ }\href
  {\doibase 10.1103/PhysRevD.107.072007} {\bibfield  {journal} {\bibinfo
  {journal} {Phys. Rev. D}\ }\textbf {\bibinfo {volume} {107}},\ \bibinfo
  {pages} {072007} (\bibinfo {year} {2023})},\ \Eprint
  {http://arxiv.org/abs/2301.09721} {arXiv:2301.09721 [hep-ex]} \BibitemShut
  {NoStop}%
\bibitem [{\citenamefont {Chang}\ \emph {et~al.}(2022)\citenamefont {Chang}
  \emph {et~al.}}]{TASEH:2022vvu}%
  \BibitemOpen
  \bibfield  {author} {\bibinfo {author} {\bibfnamefont {H.}~\bibnamefont
  {Chang}} \emph {et~al.} (\bibinfo {collaboration} {TASEH}),\ }\href {\doibase
  10.1103/PhysRevLett.129.111802} {\bibfield  {journal} {\bibinfo  {journal}
  {Phys. Rev. Lett.}\ }\textbf {\bibinfo {volume} {129}},\ \bibinfo {pages}
  {111802} (\bibinfo {year} {2022})},\ \Eprint
  {http://arxiv.org/abs/2205.05574} {arXiv:2205.05574 [hep-ex]} \BibitemShut
  {NoStop}%
\bibitem [{\citenamefont {Quiskamp}\ \emph {et~al.}(2022)\citenamefont
  {Quiskamp}, \citenamefont {McAllister}, \citenamefont {Altin}, \citenamefont
  {Ivanov}, \citenamefont {Goryachev},\ and\ \citenamefont
  {Tobar}}]{Quiskamp:2022pks}%
  \BibitemOpen
  \bibfield  {author} {\bibinfo {author} {\bibfnamefont {A.~P.}\ \bibnamefont
  {Quiskamp}}, \bibinfo {author} {\bibfnamefont {B.~T.}\ \bibnamefont
  {McAllister}}, \bibinfo {author} {\bibfnamefont {P.}~\bibnamefont {Altin}},
  \bibinfo {author} {\bibfnamefont {E.~N.}\ \bibnamefont {Ivanov}}, \bibinfo
  {author} {\bibfnamefont {M.}~\bibnamefont {Goryachev}}, \ and\ \bibinfo
  {author} {\bibfnamefont {M.~E.}\ \bibnamefont {Tobar}},\ }\href {\doibase
  10.1126/sciadv.abq3765} {\bibfield  {journal} {\bibinfo  {journal} {Sci.
  Adv.}\ }\textbf {\bibinfo {volume} {8}},\ \bibinfo {pages} {abq3765}
  (\bibinfo {year} {2022})},\ \Eprint {http://arxiv.org/abs/2203.12152}
  {arXiv:2203.12152 [hep-ex]} \BibitemShut {NoStop}%
\bibitem [{\citenamefont {Quiskamp}\ \emph {et~al.}(2024)\citenamefont
  {Quiskamp}, \citenamefont {McAllister}, \citenamefont {Altin}, \citenamefont
  {Ivanov}, \citenamefont {Goryachev},\ and\ \citenamefont
  {Tobar}}]{Quiskamp:2023ehr}%
  \BibitemOpen
  \bibfield  {author} {\bibinfo {author} {\bibfnamefont {A.}~\bibnamefont
  {Quiskamp}}, \bibinfo {author} {\bibfnamefont {B.~T.}\ \bibnamefont
  {McAllister}}, \bibinfo {author} {\bibfnamefont {P.}~\bibnamefont {Altin}},
  \bibinfo {author} {\bibfnamefont {E.~N.}\ \bibnamefont {Ivanov}}, \bibinfo
  {author} {\bibfnamefont {M.}~\bibnamefont {Goryachev}}, \ and\ \bibinfo
  {author} {\bibfnamefont {M.~E.}\ \bibnamefont {Tobar}},\ }\href {\doibase
  10.1103/PhysRevLett.132.031601} {\bibfield  {journal} {\bibinfo  {journal}
  {Phys. Rev. Lett.}\ }\textbf {\bibinfo {volume} {132}},\ \bibinfo {pages}
  {031601} (\bibinfo {year} {2024})},\ \Eprint
  {http://arxiv.org/abs/2310.00904} {arXiv:2310.00904 [hep-ex]} \BibitemShut
  {NoStop}%
\bibitem [{\citenamefont {Lee}\ \emph {et~al.}(2020)\citenamefont {Lee},
  \citenamefont {Ahn}, \citenamefont {Choi}, \citenamefont {Ko},\ and\
  \citenamefont {Semertzidis}}]{Lee:2020cfj}%
  \BibitemOpen
  \bibfield  {author} {\bibinfo {author} {\bibfnamefont {S.}~\bibnamefont
  {Lee}}, \bibinfo {author} {\bibfnamefont {S.}~\bibnamefont {Ahn}}, \bibinfo
  {author} {\bibfnamefont {J.}~\bibnamefont {Choi}}, \bibinfo {author}
  {\bibfnamefont {B.~R.}\ \bibnamefont {Ko}}, \ and\ \bibinfo {author}
  {\bibfnamefont {Y.~K.}\ \bibnamefont {Semertzidis}},\ }\href {\doibase
  10.1103/PhysRevLett.124.101802} {\bibfield  {journal} {\bibinfo  {journal}
  {Phys. Rev. Lett.}\ }\textbf {\bibinfo {volume} {124}},\ \bibinfo {pages}
  {101802} (\bibinfo {year} {2020})},\ \Eprint
  {http://arxiv.org/abs/2001.05102} {arXiv:2001.05102 [hep-ex]} \BibitemShut
  {NoStop}%
\bibitem [{\citenamefont {Jeong}\ \emph {et~al.}(2020)\citenamefont {Jeong},
  \citenamefont {Youn}, \citenamefont {Bae}, \citenamefont {Kim}, \citenamefont
  {Seong}, \citenamefont {Kim},\ and\ \citenamefont
  {Semertzidis}}]{Jeong:2020cwz}%
  \BibitemOpen
  \bibfield  {author} {\bibinfo {author} {\bibfnamefont {J.}~\bibnamefont
  {Jeong}}, \bibinfo {author} {\bibfnamefont {S.}~\bibnamefont {Youn}},
  \bibinfo {author} {\bibfnamefont {S.}~\bibnamefont {Bae}}, \bibinfo {author}
  {\bibfnamefont {J.}~\bibnamefont {Kim}}, \bibinfo {author} {\bibfnamefont
  {T.}~\bibnamefont {Seong}}, \bibinfo {author} {\bibfnamefont {J.~E.}\
  \bibnamefont {Kim}}, \ and\ \bibinfo {author} {\bibfnamefont {Y.~K.}\
  \bibnamefont {Semertzidis}},\ }\href {\doibase
  10.1103/PhysRevLett.125.221302} {\bibfield  {journal} {\bibinfo  {journal}
  {Phys. Rev. Lett.}\ }\textbf {\bibinfo {volume} {125}},\ \bibinfo {pages}
  {221302} (\bibinfo {year} {2020})},\ \Eprint
  {http://arxiv.org/abs/2008.10141} {arXiv:2008.10141 [hep-ex]} \BibitemShut
  {NoStop}%
\bibitem [{\citenamefont {Kwon}\ \emph {et~al.}(2021)\citenamefont {Kwon} \emph
  {et~al.}}]{CAPP:2020utb}%
  \BibitemOpen
  \bibfield  {author} {\bibinfo {author} {\bibfnamefont {O.}~\bibnamefont
  {Kwon}} \emph {et~al.} (\bibinfo {collaboration} {CAPP}),\ }\href {\doibase
  10.1103/PhysRevLett.126.191802} {\bibfield  {journal} {\bibinfo  {journal}
  {Phys. Rev. Lett.}\ }\textbf {\bibinfo {volume} {126}},\ \bibinfo {pages}
  {191802} (\bibinfo {year} {2021})},\ \Eprint
  {http://arxiv.org/abs/2012.10764} {arXiv:2012.10764 [hep-ex]} \BibitemShut
  {NoStop}%
\bibitem [{\citenamefont {Lee}\ \emph {et~al.}(2022)\citenamefont {Lee},
  \citenamefont {Yang}, \citenamefont {Yoon}, \citenamefont {Ahn},
  \citenamefont {Park}, \citenamefont {Min}, \citenamefont {Kim},\ and\
  \citenamefont {Yoo}}]{Lee:2022mnc}%
  \BibitemOpen
  \bibfield  {author} {\bibinfo {author} {\bibfnamefont {Y.}~\bibnamefont
  {Lee}}, \bibinfo {author} {\bibfnamefont {B.}~\bibnamefont {Yang}}, \bibinfo
  {author} {\bibfnamefont {H.}~\bibnamefont {Yoon}}, \bibinfo {author}
  {\bibfnamefont {M.}~\bibnamefont {Ahn}}, \bibinfo {author} {\bibfnamefont
  {H.}~\bibnamefont {Park}}, \bibinfo {author} {\bibfnamefont {B.}~\bibnamefont
  {Min}}, \bibinfo {author} {\bibfnamefont {D.}~\bibnamefont {Kim}}, \ and\
  \bibinfo {author} {\bibfnamefont {J.}~\bibnamefont {Yoo}},\ }\href {\doibase
  10.1103/PhysRevLett.128.241805} {\bibfield  {journal} {\bibinfo  {journal}
  {Phys. Rev. Lett.}\ }\textbf {\bibinfo {volume} {128}},\ \bibinfo {pages}
  {241805} (\bibinfo {year} {2022})},\ \Eprint
  {http://arxiv.org/abs/2206.08845} {arXiv:2206.08845 [hep-ex]} \BibitemShut
  {NoStop}%
\bibitem [{\citenamefont {Kim}\ \emph {et~al.}(2023{\natexlab{a}})\citenamefont
  {Kim} \emph {et~al.}}]{Kim:2022hmg}%
  \BibitemOpen
  \bibfield  {author} {\bibinfo {author} {\bibfnamefont {J.}~\bibnamefont
  {Kim}} \emph {et~al.},\ }\href {\doibase 10.1103/PhysRevLett.130.091602}
  {\bibfield  {journal} {\bibinfo  {journal} {Phys. Rev. Lett.}\ }\textbf
  {\bibinfo {volume} {130}},\ \bibinfo {pages} {091602} (\bibinfo {year}
  {2023}{\natexlab{a}})},\ \Eprint {http://arxiv.org/abs/2207.13597}
  {arXiv:2207.13597 [hep-ex]} \BibitemShut {NoStop}%
\bibitem [{\citenamefont {Yi}\ \emph {et~al.}(2023)\citenamefont {Yi} \emph
  {et~al.}}]{Yi:2022fmn}%
  \BibitemOpen
  \bibfield  {author} {\bibinfo {author} {\bibfnamefont {A.~K.}\ \bibnamefont
  {Yi}} \emph {et~al.},\ }\href {\doibase 10.1103/PhysRevLett.130.071002}
  {\bibfield  {journal} {\bibinfo  {journal} {Phys. Rev. Lett.}\ }\textbf
  {\bibinfo {volume} {130}},\ \bibinfo {pages} {071002} (\bibinfo {year}
  {2023})},\ \Eprint {http://arxiv.org/abs/2210.10961} {arXiv:2210.10961
  [hep-ex]} \BibitemShut {NoStop}%
\bibitem [{\citenamefont {Yang}\ \emph {et~al.}(2023)\citenamefont {Yang},
  \citenamefont {Yoon}, \citenamefont {Ahn}, \citenamefont {Lee},\ and\
  \citenamefont {Yoo}}]{Yang:2023yry}%
  \BibitemOpen
  \bibfield  {author} {\bibinfo {author} {\bibfnamefont {B.}~\bibnamefont
  {Yang}}, \bibinfo {author} {\bibfnamefont {H.}~\bibnamefont {Yoon}}, \bibinfo
  {author} {\bibfnamefont {M.}~\bibnamefont {Ahn}}, \bibinfo {author}
  {\bibfnamefont {Y.}~\bibnamefont {Lee}}, \ and\ \bibinfo {author}
  {\bibfnamefont {J.}~\bibnamefont {Yoo}},\ }\href {\doibase
  10.1103/PhysRevLett.131.081801} {\bibfield  {journal} {\bibinfo  {journal}
  {Phys. Rev. Lett.}\ }\textbf {\bibinfo {volume} {131}},\ \bibinfo {pages}
  {081801} (\bibinfo {year} {2023})},\ \Eprint
  {http://arxiv.org/abs/2308.09077} {arXiv:2308.09077 [hep-ex]} \BibitemShut
  {NoStop}%
\bibitem [{\citenamefont {Kim}\ \emph {et~al.}(2023{\natexlab{b}})\citenamefont
  {Kim} \emph {et~al.}}]{Kim:2023vpo}%
  \BibitemOpen
  \bibfield  {author} {\bibinfo {author} {\bibfnamefont {Y.}~\bibnamefont
  {Kim}} \emph {et~al.},\ }\href@noop {} {\  (\bibinfo {year}
  {2023}{\natexlab{b}})},\ \Eprint {http://arxiv.org/abs/2312.11003}
  {arXiv:2312.11003 [hep-ex]} \BibitemShut {NoStop}%
\bibitem [{\citenamefont {Adair}\ \emph {et~al.}(2022)\citenamefont {Adair}
  \emph {et~al.}}]{Adair:2022rtw}%
  \BibitemOpen
  \bibfield  {author} {\bibinfo {author} {\bibfnamefont {C.~M.}\ \bibnamefont
  {Adair}} \emph {et~al.},\ }\href {\doibase 10.1038/s41467-022-33913-6}
  {\bibfield  {journal} {\bibinfo  {journal} {Nature Commun.}\ }\textbf
  {\bibinfo {volume} {13}},\ \bibinfo {pages} {6180} (\bibinfo {year}
  {2022})},\ \Eprint {http://arxiv.org/abs/2211.02902} {arXiv:2211.02902
  [hep-ex]} \BibitemShut {NoStop}%
\bibitem [{\citenamefont {O'Hare}(2020)}]{AxionLimits}%
  \BibitemOpen
  \bibfield  {author} {\bibinfo {author} {\bibfnamefont {C.}~\bibnamefont
  {O'Hare}},\ }\href {\doibase 10.5281/zenodo.3932430} {\enquote {\bibinfo
  {title} {cajohare/axionlimits: Axionlimits},}\ }\bibinfo {howpublished}
  {\url{https://cajohare.github.io/AxionLimits/}} (\bibinfo {year}
  {2020})\BibitemShut {NoStop}%
\bibitem [{\citenamefont {Raffelt}(2008)}]{Raffelt:2006cw}%
  \BibitemOpen
  \bibfield  {author} {\bibinfo {author} {\bibfnamefont {G.~G.}\ \bibnamefont
  {Raffelt}},\ }\href {\doibase 10.1007/978-3-540-73518-2_3} {\bibfield
  {journal} {\bibinfo  {journal} {Lect. Notes Phys.}\ }\textbf {\bibinfo
  {volume} {741}},\ \bibinfo {pages} {51} (\bibinfo {year} {2008})},\ \Eprint
  {http://arxiv.org/abs/hep-ph/0611350} {arXiv:hep-ph/0611350} \BibitemShut
  {NoStop}%
\bibitem [{\citenamefont {Caputo}\ and\ \citenamefont
  {Raffelt}(2024)}]{Caputo:2024oqc}%
  \BibitemOpen
  \bibfield  {author} {\bibinfo {author} {\bibfnamefont {A.}~\bibnamefont
  {Caputo}}\ and\ \bibinfo {author} {\bibfnamefont {G.}~\bibnamefont
  {Raffelt}},\ }in\ \href@noop {} {\emph {\bibinfo {booktitle} {{1st Training
  School of the COST Action COSMIC WISPers (CA21106)}}}}\ (\bibinfo {year}
  {2024})\ \Eprint {http://arxiv.org/abs/2401.13728} {arXiv:2401.13728
  [hep-ph]} \BibitemShut {NoStop}%
\bibitem [{\citenamefont {Hook}\ and\ \citenamefont
  {Huang}(2018)}]{Hook:2017psm}%
  \BibitemOpen
  \bibfield  {author} {\bibinfo {author} {\bibfnamefont {A.}~\bibnamefont
  {Hook}}\ and\ \bibinfo {author} {\bibfnamefont {J.}~\bibnamefont {Huang}},\
  }\href {\doibase 10.1007/JHEP06(2018)036} {\bibfield  {journal} {\bibinfo
  {journal} {JHEP}\ }\textbf {\bibinfo {volume} {06}},\ \bibinfo {pages} {036}
  (\bibinfo {year} {2018})},\ \Eprint {http://arxiv.org/abs/1708.08464}
  {arXiv:1708.08464 [hep-ph]} \BibitemShut {NoStop}%
\bibitem [{\citenamefont {Di~Luzio}\ \emph
  {et~al.}(2021{\natexlab{b}})\citenamefont {Di~Luzio}, \citenamefont {Gavela},
  \citenamefont {Quilez},\ and\ \citenamefont {Ringwald}}]{DiLuzio:2021pxd}%
  \BibitemOpen
  \bibfield  {author} {\bibinfo {author} {\bibfnamefont {L.}~\bibnamefont
  {Di~Luzio}}, \bibinfo {author} {\bibfnamefont {B.}~\bibnamefont {Gavela}},
  \bibinfo {author} {\bibfnamefont {P.}~\bibnamefont {Quilez}}, \ and\ \bibinfo
  {author} {\bibfnamefont {A.}~\bibnamefont {Ringwald}},\ }\href {\doibase
  10.1007/JHEP05(2021)184} {\bibfield  {journal} {\bibinfo  {journal} {JHEP}\
  }\textbf {\bibinfo {volume} {05}},\ \bibinfo {pages} {184} (\bibinfo {year}
  {2021}{\natexlab{b}})},\ \Eprint {http://arxiv.org/abs/2102.00012}
  {arXiv:2102.00012 [hep-ph]} \BibitemShut {NoStop}%
\bibitem [{\citenamefont {Abel}\ \emph {et~al.}(2017)\citenamefont {Abel} \emph
  {et~al.}}]{Abel:2017rtm}%
  \BibitemOpen
  \bibfield  {author} {\bibinfo {author} {\bibfnamefont {C.}~\bibnamefont
  {Abel}} \emph {et~al.},\ }\href {\doibase 10.1103/PhysRevX.7.041034}
  {\bibfield  {journal} {\bibinfo  {journal} {Phys. Rev. X}\ }\textbf {\bibinfo
  {volume} {7}},\ \bibinfo {pages} {041034} (\bibinfo {year} {2017})},\ \Eprint
  {http://arxiv.org/abs/1708.06367} {arXiv:1708.06367 [hep-ph]} \BibitemShut
  {NoStop}%
\bibitem [{\citenamefont {Schulthess}\ \emph {et~al.}(2022)\citenamefont
  {Schulthess} \emph {et~al.}}]{Schulthess:2022pbp}%
  \BibitemOpen
  \bibfield  {author} {\bibinfo {author} {\bibfnamefont {I.}~\bibnamefont
  {Schulthess}} \emph {et~al.},\ }\href {\doibase
  10.1103/PhysRevLett.129.191801} {\bibfield  {journal} {\bibinfo  {journal}
  {Phys. Rev. Lett.}\ }\textbf {\bibinfo {volume} {129}},\ \bibinfo {pages}
  {191801} (\bibinfo {year} {2022})},\ \Eprint
  {http://arxiv.org/abs/2204.01454} {arXiv:2204.01454 [hep-ex]} \BibitemShut
  {NoStop}%
\bibitem [{\citenamefont {Balkin}\ \emph {et~al.}(2022)\citenamefont {Balkin},
  \citenamefont {Serra}, \citenamefont {Springmann}, \citenamefont {Stelzl},\
  and\ \citenamefont {Weiler}}]{Balkin:2022qer}%
  \BibitemOpen
  \bibfield  {author} {\bibinfo {author} {\bibfnamefont {R.}~\bibnamefont
  {Balkin}}, \bibinfo {author} {\bibfnamefont {J.}~\bibnamefont {Serra}},
  \bibinfo {author} {\bibfnamefont {K.}~\bibnamefont {Springmann}}, \bibinfo
  {author} {\bibfnamefont {S.}~\bibnamefont {Stelzl}}, \ and\ \bibinfo {author}
  {\bibfnamefont {A.}~\bibnamefont {Weiler}},\ }\href@noop {} {\  (\bibinfo
  {year} {2022})},\ \Eprint {http://arxiv.org/abs/2211.02661} {arXiv:2211.02661
  [hep-ph]} \BibitemShut {NoStop}%
\bibitem [{\citenamefont {Budker}\ \emph {et~al.}(2014)\citenamefont {Budker},
  \citenamefont {Graham}, \citenamefont {Ledbetter}, \citenamefont
  {Rajendran},\ and\ \citenamefont {Sushkov}}]{Budker:2013hfa}%
  \BibitemOpen
  \bibfield  {author} {\bibinfo {author} {\bibfnamefont {D.}~\bibnamefont
  {Budker}}, \bibinfo {author} {\bibfnamefont {P.~W.}\ \bibnamefont {Graham}},
  \bibinfo {author} {\bibfnamefont {M.}~\bibnamefont {Ledbetter}}, \bibinfo
  {author} {\bibfnamefont {S.}~\bibnamefont {Rajendran}}, \ and\ \bibinfo
  {author} {\bibfnamefont {A.}~\bibnamefont {Sushkov}},\ }\href {\doibase
  10.1103/PhysRevX.4.021030} {\bibfield  {journal} {\bibinfo  {journal} {Phys.
  Rev. X}\ }\textbf {\bibinfo {volume} {4}},\ \bibinfo {pages} {021030}
  (\bibinfo {year} {2014})},\ \Eprint {http://arxiv.org/abs/1306.6089}
  {arXiv:1306.6089 [hep-ph]} \BibitemShut {NoStop}%
\bibitem [{\citenamefont {Jackson~Kimball}\ \emph {et~al.}(2020)\citenamefont
  {Jackson~Kimball} \emph {et~al.}}]{JacksonKimball:2017elr}%
  \BibitemOpen
  \bibfield  {author} {\bibinfo {author} {\bibfnamefont {D.~F.}\ \bibnamefont
  {Jackson~Kimball}} \emph {et~al.},\ }\href {\doibase
  10.1007/978-3-030-43761-9_13} {\bibfield  {journal} {\bibinfo  {journal}
  {Springer Proc. Phys.}\ }\textbf {\bibinfo {volume} {245}},\ \bibinfo {pages}
  {105} (\bibinfo {year} {2020})},\ \Eprint {http://arxiv.org/abs/1711.08999}
  {arXiv:1711.08999 [physics.ins-det]} \BibitemShut {NoStop}%
\bibitem [{\citenamefont {Workman}\ \emph {et~al.}(2022)\citenamefont {Workman}
  \emph {et~al.}}]{ParticleDataGroup:2022pth}%
  \BibitemOpen
  \bibfield  {author} {\bibinfo {author} {\bibfnamefont {R.~L.}\ \bibnamefont
  {Workman}} \emph {et~al.} (\bibinfo {collaboration} {Particle Data Group}),\
  }\href {\doibase 10.1093/ptep/ptac097} {\bibfield  {journal} {\bibinfo
  {journal} {PTEP}\ }\textbf {\bibinfo {volume} {2022}},\ \bibinfo {pages}
  {083C01} (\bibinfo {year} {2022})}\BibitemShut {NoStop}%
\bibitem [{\citenamefont {Fox}\ \emph {et~al.}(2004)\citenamefont {Fox},
  \citenamefont {Pierce},\ and\ \citenamefont {Thomas}}]{Fox:2004kb}%
  \BibitemOpen
  \bibfield  {author} {\bibinfo {author} {\bibfnamefont {P.}~\bibnamefont
  {Fox}}, \bibinfo {author} {\bibfnamefont {A.}~\bibnamefont {Pierce}}, \ and\
  \bibinfo {author} {\bibfnamefont {S.~D.}\ \bibnamefont {Thomas}},\
  }\href@noop {} {\  (\bibinfo {year} {2004})},\ \Eprint
  {http://arxiv.org/abs/hep-th/0409059} {arXiv:hep-th/0409059} \BibitemShut
  {NoStop}%
\bibitem [{\citenamefont {Beltran}\ \emph {et~al.}(2007)\citenamefont
  {Beltran}, \citenamefont {Garcia-Bellido},\ and\ \citenamefont
  {Lesgourgues}}]{Beltran:2006sq}%
  \BibitemOpen
  \bibfield  {author} {\bibinfo {author} {\bibfnamefont {M.}~\bibnamefont
  {Beltran}}, \bibinfo {author} {\bibfnamefont {J.}~\bibnamefont
  {Garcia-Bellido}}, \ and\ \bibinfo {author} {\bibfnamefont {J.}~\bibnamefont
  {Lesgourgues}},\ }\href {\doibase 10.1103/PhysRevD.75.103507} {\bibfield
  {journal} {\bibinfo  {journal} {Phys. Rev. D}\ }\textbf {\bibinfo {volume}
  {75}},\ \bibinfo {pages} {103507} (\bibinfo {year} {2007})},\ \Eprint
  {http://arxiv.org/abs/hep-ph/0606107} {arXiv:hep-ph/0606107} \BibitemShut
  {NoStop}%
\bibitem [{\citenamefont {Acharya}\ \emph {et~al.}(2010)\citenamefont
  {Acharya}, \citenamefont {Bobkov},\ and\ \citenamefont
  {Kumar}}]{Acharya:2010zx}%
  \BibitemOpen
  \bibfield  {author} {\bibinfo {author} {\bibfnamefont {B.~S.}\ \bibnamefont
  {Acharya}}, \bibinfo {author} {\bibfnamefont {K.}~\bibnamefont {Bobkov}}, \
  and\ \bibinfo {author} {\bibfnamefont {P.}~\bibnamefont {Kumar}},\ }\href
  {\doibase 10.1007/JHEP11(2010)105} {\bibfield  {journal} {\bibinfo  {journal}
  {JHEP}\ }\textbf {\bibinfo {volume} {11}},\ \bibinfo {pages} {105} (\bibinfo
  {year} {2010})},\ \Eprint {http://arxiv.org/abs/1004.5138} {arXiv:1004.5138
  [hep-th]} \BibitemShut {NoStop}%
\bibitem [{\citenamefont {Marsh}\ \emph {et~al.}(2014)\citenamefont {Marsh},
  \citenamefont {Grin}, \citenamefont {Hlozek},\ and\ \citenamefont
  {Ferreira}}]{Marsh:2014qoa}%
  \BibitemOpen
  \bibfield  {author} {\bibinfo {author} {\bibfnamefont {D.~J.~E.}\
  \bibnamefont {Marsh}}, \bibinfo {author} {\bibfnamefont {D.}~\bibnamefont
  {Grin}}, \bibinfo {author} {\bibfnamefont {R.}~\bibnamefont {Hlozek}}, \ and\
  \bibinfo {author} {\bibfnamefont {P.~G.}\ \bibnamefont {Ferreira}},\ }\href
  {\doibase 10.1103/PhysRevLett.113.011801} {\bibfield  {journal} {\bibinfo
  {journal} {Phys. Rev. Lett.}\ }\textbf {\bibinfo {volume} {113}},\ \bibinfo
  {pages} {011801} (\bibinfo {year} {2014})},\ \Eprint
  {http://arxiv.org/abs/1403.4216} {arXiv:1403.4216 [astro-ph.CO]} \BibitemShut
  {NoStop}%
\bibitem [{\citenamefont {Chen}\ \emph {et~al.}(2023)\citenamefont {Chen},
  \citenamefont {Fan},\ and\ \citenamefont {Li}}]{Chen:2023txq}%
  \BibitemOpen
  \bibfield  {author} {\bibinfo {author} {\bibfnamefont {X.}~\bibnamefont
  {Chen}}, \bibinfo {author} {\bibfnamefont {J.}~\bibnamefont {Fan}}, \ and\
  \bibinfo {author} {\bibfnamefont {L.}~\bibnamefont {Li}},\ }\href {\doibase
  10.1007/JHEP12(2023)197} {\bibfield  {journal} {\bibinfo  {journal} {JHEP}\
  }\textbf {\bibinfo {volume} {12}},\ \bibinfo {pages} {197} (\bibinfo {year}
  {2023})},\ \Eprint {http://arxiv.org/abs/2303.03406} {arXiv:2303.03406
  [hep-ph]} \BibitemShut {NoStop}%
\bibitem [{\citenamefont {Rogers}\ and\ \citenamefont
  {Peiris}(2021)}]{Rogers:2020ltq}%
  \BibitemOpen
  \bibfield  {author} {\bibinfo {author} {\bibfnamefont {K.~K.}\ \bibnamefont
  {Rogers}}\ and\ \bibinfo {author} {\bibfnamefont {H.~V.}\ \bibnamefont
  {Peiris}},\ }\href {\doibase 10.1103/PhysRevLett.126.071302} {\bibfield
  {journal} {\bibinfo  {journal} {Phys. Rev. Lett.}\ }\textbf {\bibinfo
  {volume} {126}},\ \bibinfo {pages} {071302} (\bibinfo {year} {2021})},\
  \Eprint {http://arxiv.org/abs/2007.12705} {arXiv:2007.12705 [astro-ph.CO]}
  \BibitemShut {NoStop}%
\bibitem [{\citenamefont {Afzal}\ \emph {et~al.}(2023)\citenamefont {Afzal}
  \emph {et~al.}}]{NANOGrav:2023hvm}%
  \BibitemOpen
  \bibfield  {author} {\bibinfo {author} {\bibfnamefont {A.}~\bibnamefont
  {Afzal}} \emph {et~al.} (\bibinfo {collaboration} {NANOGrav}),\ }\href
  {\doibase 10.3847/2041-8213/acdc91} {\bibfield  {journal} {\bibinfo
  {journal} {Astrophys. J. Lett.}\ }\textbf {\bibinfo {volume} {951}},\
  \bibinfo {pages} {L11} (\bibinfo {year} {2023})},\ \Eprint
  {http://arxiv.org/abs/2306.16219} {arXiv:2306.16219 [astro-ph.HE]}
  \BibitemShut {NoStop}%
\bibitem [{\citenamefont {Marsh}\ and\ \citenamefont
  {Niemeyer}(2019)}]{Marsh:2018zyw}%
  \BibitemOpen
  \bibfield  {author} {\bibinfo {author} {\bibfnamefont {D.~J.~E.}\
  \bibnamefont {Marsh}}\ and\ \bibinfo {author} {\bibfnamefont {J.~C.}\
  \bibnamefont {Niemeyer}},\ }\href {\doibase 10.1103/PhysRevLett.123.051103}
  {\bibfield  {journal} {\bibinfo  {journal} {Phys. Rev. Lett.}\ }\textbf
  {\bibinfo {volume} {123}},\ \bibinfo {pages} {051103} (\bibinfo {year}
  {2019})},\ \Eprint {http://arxiv.org/abs/1810.08543} {arXiv:1810.08543
  [astro-ph.CO]} \BibitemShut {NoStop}%
\bibitem [{\citenamefont {Nadler}\ \emph {et~al.}(2021)\citenamefont {Nadler}
  \emph {et~al.}}]{DES:2020fxi}%
  \BibitemOpen
  \bibfield  {author} {\bibinfo {author} {\bibfnamefont {E.~O.}\ \bibnamefont
  {Nadler}} \emph {et~al.} (\bibinfo {collaboration} {DES}),\ }\href {\doibase
  10.1103/PhysRevLett.126.091101} {\bibfield  {journal} {\bibinfo  {journal}
  {Phys. Rev. Lett.}\ }\textbf {\bibinfo {volume} {126}},\ \bibinfo {pages}
  {091101} (\bibinfo {year} {2021})},\ \Eprint
  {http://arxiv.org/abs/2008.00022} {arXiv:2008.00022 [astro-ph.CO]}
  \BibitemShut {NoStop}%
\bibitem [{\citenamefont {Arvanitaki}\ and\ \citenamefont
  {Dubovsky}(2011)}]{Arvanitaki:2010sy}%
  \BibitemOpen
  \bibfield  {author} {\bibinfo {author} {\bibfnamefont {A.}~\bibnamefont
  {Arvanitaki}}\ and\ \bibinfo {author} {\bibfnamefont {S.}~\bibnamefont
  {Dubovsky}},\ }\href {\doibase 10.1103/PhysRevD.83.044026} {\bibfield
  {journal} {\bibinfo  {journal} {Phys. Rev. D}\ }\textbf {\bibinfo {volume}
  {83}},\ \bibinfo {pages} {044026} (\bibinfo {year} {2011})},\ \Eprint
  {http://arxiv.org/abs/1004.3558} {arXiv:1004.3558 [hep-th]} \BibitemShut
  {NoStop}%
\bibitem [{\citenamefont {Mehta}\ \emph {et~al.}(2020)\citenamefont {Mehta},
  \citenamefont {Demirtas}, \citenamefont {Long}, \citenamefont {Marsh},
  \citenamefont {Mcallister},\ and\ \citenamefont {Stott}}]{Mehta:2020kwu}%
  \BibitemOpen
  \bibfield  {author} {\bibinfo {author} {\bibfnamefont {V.~M.}\ \bibnamefont
  {Mehta}}, \bibinfo {author} {\bibfnamefont {M.}~\bibnamefont {Demirtas}},
  \bibinfo {author} {\bibfnamefont {C.}~\bibnamefont {Long}}, \bibinfo {author}
  {\bibfnamefont {D.~J.~E.}\ \bibnamefont {Marsh}}, \bibinfo {author}
  {\bibfnamefont {L.}~\bibnamefont {Mcallister}}, \ and\ \bibinfo {author}
  {\bibfnamefont {M.~J.}\ \bibnamefont {Stott}},\ }\href@noop {} {\  (\bibinfo
  {year} {2020})},\ \Eprint {http://arxiv.org/abs/2011.08693} {arXiv:2011.08693
  [hep-th]} \BibitemShut {NoStop}%
\bibitem [{\citenamefont {Michimura}\ \emph {et~al.}(2020)\citenamefont
  {Michimura}, \citenamefont {Oshima}, \citenamefont {Watanabe}, \citenamefont
  {Kawasaki}, \citenamefont {Takeda}, \citenamefont {Ando}, \citenamefont
  {Nagano}, \citenamefont {Obata},\ and\ \citenamefont
  {Fujita}}]{Michimura:2019qxr}%
  \BibitemOpen
  \bibfield  {author} {\bibinfo {author} {\bibfnamefont {Y.}~\bibnamefont
  {Michimura}}, \bibinfo {author} {\bibfnamefont {Y.}~\bibnamefont {Oshima}},
  \bibinfo {author} {\bibfnamefont {T.}~\bibnamefont {Watanabe}}, \bibinfo
  {author} {\bibfnamefont {T.}~\bibnamefont {Kawasaki}}, \bibinfo {author}
  {\bibfnamefont {H.}~\bibnamefont {Takeda}}, \bibinfo {author} {\bibfnamefont
  {M.}~\bibnamefont {Ando}}, \bibinfo {author} {\bibfnamefont {K.}~\bibnamefont
  {Nagano}}, \bibinfo {author} {\bibfnamefont {I.}~\bibnamefont {Obata}}, \
  and\ \bibinfo {author} {\bibfnamefont {T.}~\bibnamefont {Fujita}},\ }\href
  {\doibase 10.1088/1742-6596/1468/1/012032} {\bibfield  {journal} {\bibinfo
  {journal} {J. Phys. Conf. Ser.}\ }\textbf {\bibinfo {volume} {1468}},\
  \bibinfo {pages} {012032} (\bibinfo {year} {2020})},\ \Eprint
  {http://arxiv.org/abs/1911.05196} {arXiv:1911.05196 [physics.ins-det]}
  \BibitemShut {NoStop}%
\bibitem [{\citenamefont {Berlin}\ \emph {et~al.}(2021)\citenamefont {Berlin},
  \citenamefont {D'Agnolo}, \citenamefont {Ellis},\ and\ \citenamefont
  {Zhou}}]{Berlin:2020vrk}%
  \BibitemOpen
  \bibfield  {author} {\bibinfo {author} {\bibfnamefont {A.}~\bibnamefont
  {Berlin}}, \bibinfo {author} {\bibfnamefont {R.~T.}\ \bibnamefont
  {D'Agnolo}}, \bibinfo {author} {\bibfnamefont {S.~A.~R.}\ \bibnamefont
  {Ellis}}, \ and\ \bibinfo {author} {\bibfnamefont {K.}~\bibnamefont {Zhou}},\
  }\href {\doibase 10.1103/PhysRevD.104.L111701} {\bibfield  {journal}
  {\bibinfo  {journal} {Phys. Rev. D}\ }\textbf {\bibinfo {volume} {104}},\
  \bibinfo {pages} {L111701} (\bibinfo {year} {2021})},\ \Eprint
  {http://arxiv.org/abs/2007.15656} {arXiv:2007.15656 [hep-ph]} \BibitemShut
  {NoStop}%
\bibitem [{\citenamefont {Brouwer}\ \emph {et~al.}(2022)\citenamefont {Brouwer}
  \emph {et~al.}}]{DMRadio:2022pkf}%
  \BibitemOpen
  \bibfield  {author} {\bibinfo {author} {\bibfnamefont {L.}~\bibnamefont
  {Brouwer}} \emph {et~al.} (\bibinfo {collaboration} {DMRadio}),\ }\href
  {\doibase 10.1103/PhysRevD.106.103008} {\bibfield  {journal} {\bibinfo
  {journal} {Phys. Rev. D}\ }\textbf {\bibinfo {volume} {106}},\ \bibinfo
  {pages} {103008} (\bibinfo {year} {2022})},\ \Eprint
  {http://arxiv.org/abs/2204.13781} {arXiv:2204.13781 [hep-ex]} \BibitemShut
  {NoStop}%
\bibitem [{\citenamefont {Bourhill}\ \emph {et~al.}(2023)\citenamefont
  {Bourhill}, \citenamefont {Paterson}, \citenamefont {Goryachev},\ and\
  \citenamefont {Tobar}}]{Bourhill:2022alm}%
  \BibitemOpen
  \bibfield  {author} {\bibinfo {author} {\bibfnamefont {J.~F.}\ \bibnamefont
  {Bourhill}}, \bibinfo {author} {\bibfnamefont {E.~C.~I.}\ \bibnamefont
  {Paterson}}, \bibinfo {author} {\bibfnamefont {M.}~\bibnamefont {Goryachev}},
  \ and\ \bibinfo {author} {\bibfnamefont {M.~E.}\ \bibnamefont {Tobar}},\
  }\href {\doibase 10.1103/PhysRevD.108.052014} {\bibfield  {journal} {\bibinfo
   {journal} {Phys. Rev. D}\ }\textbf {\bibinfo {volume} {108}},\ \bibinfo
  {pages} {052014} (\bibinfo {year} {2023})},\ \Eprint
  {http://arxiv.org/abs/2208.01640} {arXiv:2208.01640 [hep-ph]} \BibitemShut
  {NoStop}%
\bibitem [{\citenamefont {Ismail}\ \emph {et~al.}()\citenamefont {Ismail},
  \citenamefont {Lee},\ and\ \citenamefont {Yu}}]{Ismail:2024}%
  \BibitemOpen
  \bibfield  {author} {\bibinfo {author} {\bibfnamefont {A.}~\bibnamefont
  {Ismail}}, \bibinfo {author} {\bibfnamefont {S.}~\bibnamefont {Lee}}, \ and\
  \bibinfo {author} {\bibfnamefont {B.}~\bibnamefont {Yu}},\ }\href@noop {}
  {\bibinfo  {journal} {Axion dark matter from inflation, in preparation}\
  }\BibitemShut {NoStop}%
\bibitem [{\citenamefont {Martin}\ and\ \citenamefont
  {Yokoyama}(2008)}]{Martin:2007ue}%
  \BibitemOpen
\bibfield  {journal} {  }\bibfield  {author} {\bibinfo {author} {\bibfnamefont
  {J.}~\bibnamefont {Martin}}\ and\ \bibinfo {author} {\bibfnamefont
  {J.}~\bibnamefont {Yokoyama}},\ }\href {\doibase
  10.1088/1475-7516/2008/01/025} {\bibfield  {journal} {\bibinfo  {journal}
  {JCAP}\ }\textbf {\bibinfo {volume} {01}},\ \bibinfo {pages} {025} (\bibinfo
  {year} {2008})},\ \Eprint {http://arxiv.org/abs/0711.4307} {arXiv:0711.4307
  [astro-ph]} \BibitemShut {NoStop}%
\bibitem [{\citenamefont {Watanabe}\ \emph {et~al.}(2009)\citenamefont
  {Watanabe}, \citenamefont {Kanno},\ and\ \citenamefont
  {Soda}}]{Watanabe:2009ct}%
  \BibitemOpen
  \bibfield  {author} {\bibinfo {author} {\bibfnamefont {M.-a.}\ \bibnamefont
  {Watanabe}}, \bibinfo {author} {\bibfnamefont {S.}~\bibnamefont {Kanno}}, \
  and\ \bibinfo {author} {\bibfnamefont {J.}~\bibnamefont {Soda}},\ }\href
  {\doibase 10.1103/PhysRevLett.102.191302} {\bibfield  {journal} {\bibinfo
  {journal} {Phys. Rev. Lett.}\ }\textbf {\bibinfo {volume} {102}},\ \bibinfo
  {pages} {191302} (\bibinfo {year} {2009})},\ \Eprint
  {http://arxiv.org/abs/0902.2833} {arXiv:0902.2833 [hep-th]} \BibitemShut
  {NoStop}%
\bibitem [{\citenamefont {Namba}(2012)}]{Namba:2012gg}%
  \BibitemOpen
  \bibfield  {author} {\bibinfo {author} {\bibfnamefont {R.}~\bibnamefont
  {Namba}},\ }\href {\doibase 10.1103/PhysRevD.86.083518} {\bibfield  {journal}
  {\bibinfo  {journal} {Phys. Rev. D}\ }\textbf {\bibinfo {volume} {86}},\
  \bibinfo {pages} {083518} (\bibinfo {year} {2012})},\ \Eprint
  {http://arxiv.org/abs/1207.5547} {arXiv:1207.5547 [astro-ph.CO]} \BibitemShut
  {NoStop}%
\bibitem [{\citenamefont {Nakayama}(2019)}]{Nakayama:2019rhg}%
  \BibitemOpen
  \bibfield  {author} {\bibinfo {author} {\bibfnamefont {K.}~\bibnamefont
  {Nakayama}},\ }\href {\doibase 10.1088/1475-7516/2019/10/019} {\bibfield
  {journal} {\bibinfo  {journal} {JCAP}\ }\textbf {\bibinfo {volume} {10}},\
  \bibinfo {pages} {019} (\bibinfo {year} {2019})},\ \Eprint
  {http://arxiv.org/abs/1907.06243} {arXiv:1907.06243 [hep-ph]} \BibitemShut
  {NoStop}%
\bibitem [{\citenamefont {Nakayama}(2020)}]{Nakayama:2020rka}%
  \BibitemOpen
  \bibfield  {author} {\bibinfo {author} {\bibfnamefont {K.}~\bibnamefont
  {Nakayama}},\ }\href {\doibase 10.1088/1475-7516/2020/08/033} {\bibfield
  {journal} {\bibinfo  {journal} {JCAP}\ }\textbf {\bibinfo {volume} {08}},\
  \bibinfo {pages} {033} (\bibinfo {year} {2020})},\ \Eprint
  {http://arxiv.org/abs/2004.10036} {arXiv:2004.10036 [hep-ph]} \BibitemShut
  {NoStop}%
\bibitem [{\citenamefont {Fairbairn}\ \emph {et~al.}(2015)\citenamefont
  {Fairbairn}, \citenamefont {Hogan},\ and\ \citenamefont
  {Marsh}}]{Fairbairn:2014zta}%
  \BibitemOpen
  \bibfield  {author} {\bibinfo {author} {\bibfnamefont {M.}~\bibnamefont
  {Fairbairn}}, \bibinfo {author} {\bibfnamefont {R.}~\bibnamefont {Hogan}}, \
  and\ \bibinfo {author} {\bibfnamefont {D.~J.~E.}\ \bibnamefont {Marsh}},\
  }\href {\doibase 10.1103/PhysRevD.91.023509} {\bibfield  {journal} {\bibinfo
  {journal} {Phys. Rev. D}\ }\textbf {\bibinfo {volume} {91}},\ \bibinfo
  {pages} {023509} (\bibinfo {year} {2015})},\ \Eprint
  {http://arxiv.org/abs/1410.1752} {arXiv:1410.1752 [hep-ph]} \BibitemShut
  {NoStop}%
\bibitem [{\citenamefont {Lee}\ \emph {et~al.}(2023)\citenamefont {Lee},
  \citenamefont {Menkara}, \citenamefont {Seong},\ and\ \citenamefont
  {Song}}]{Lee:2023dtw}%
  \BibitemOpen
  \bibfield  {author} {\bibinfo {author} {\bibfnamefont {H.~M.}\ \bibnamefont
  {Lee}}, \bibinfo {author} {\bibfnamefont {A.~G.}\ \bibnamefont {Menkara}},
  \bibinfo {author} {\bibfnamefont {M.-J.}\ \bibnamefont {Seong}}, \ and\
  \bibinfo {author} {\bibfnamefont {J.-H.}\ \bibnamefont {Song}},\ }\href@noop
  {} {\  (\bibinfo {year} {2023})},\ \Eprint {http://arxiv.org/abs/2310.17710}
  {arXiv:2310.17710 [hep-ph]} \BibitemShut {NoStop}%
\bibitem [{\citenamefont {Linde}\ \emph {et~al.}(2011)\citenamefont {Linde},
  \citenamefont {Noorbala},\ and\ \citenamefont {Westphal}}]{Linde:2011nh}%
  \BibitemOpen
  \bibfield  {author} {\bibinfo {author} {\bibfnamefont {A.}~\bibnamefont
  {Linde}}, \bibinfo {author} {\bibfnamefont {M.}~\bibnamefont {Noorbala}}, \
  and\ \bibinfo {author} {\bibfnamefont {A.}~\bibnamefont {Westphal}},\ }\href
  {\doibase 10.1088/1475-7516/2011/03/013} {\bibfield  {journal} {\bibinfo
  {journal} {JCAP}\ }\textbf {\bibinfo {volume} {03}},\ \bibinfo {pages} {013}
  (\bibinfo {year} {2011})},\ \Eprint {http://arxiv.org/abs/1101.2652}
  {arXiv:1101.2652 [hep-th]} \BibitemShut {NoStop}%
\end{thebibliography}%

\end{document}